\documentclass[aps,showpacs,preprintnumbers,amsmath, amssymb]{revtex4}

\usepackage{float}
\usepackage{graphics,epsfig}
\usepackage{graphicx}
\usepackage{dcolumn}
\usepackage{bm}
\usepackage{booktabs}

\begin{document}
\baselineskip=0.8 cm
\title{{\bf Strong gravitational lensing in a rotating Kaluza-Klein black hole with squashed horizons}}

\author{Liyong Ji, Songbai Chen\footnote{csb3752@hunnu.edu.cn}, Jiliang Jing
\footnote{jljing@hunnu.edu.cn}}

\affiliation{Institute of Physics and Department of Physics, Hunan
Normal University,  Changsha, Hunan 410081, People's Republic of
China \\ Key Laboratory of Low Dimensional Quantum Structures \\
and Quantum Control of Ministry of Education, Hunan Normal
University, Changsha, Hunan 410081, People's Republic of China}

\vspace*{0.2cm}
\begin{abstract}
\baselineskip=0.6 cm
\begin{center}
{\bf Abstract}
\end{center}

We have investigated the strong gravitational lensing in a rotating squashed Kaluza-Klein (KK) black hole spacetime. Our result show that the strong gravitational lensings in the rotating squashed KK black hole spacetime have some distinct behaviors from those
in the backgrounds of the four-dimensional Kerr black hole and of the squashed KK G\"{o}del black hole. In the rotating squashed KK black hole spacetime, the marginally circular photon radius $\rho_{ps}$ , the coefficient $\bar{a}$, $\bar{b}$, the deflection angle $\alpha(\theta)$ in the $\phi$
direction and the corresponding observational variables are independent of whether the photon goes with or against
the rotation of the background, which is different with those in the usual four-dimensional Kerr black hole spacetime. Moreover, we also find that with the increase of the scale of extra dimension $\rho_0$,  the marginally circular photon radius $\rho_{ps}$ and the angular position of the relativistic images $\theta_\infty$  first decreases and then increases in the rotating squashed KK black hole for fixed rotation parameter $b$, but in the squashed
KK G\"{o}del black hole they increase for the smaller global rotation parameter $j$ and decrease for the larger one. In the extremely squashed case $\rho_0=0$, the coefficient $\bar{a}$ in the rotating squashed KK black hole increases monotonously with the rotation parameter, but in the squashed
KK G\"{o}del black hole it is a constant and independent
of the global rotation of the G\"{o}del Universe. These information
could help us to understand further the effects of the rotation parameter and the scale of extra dimension on the strong gravitational lensing in the black hole spacetimes.
\end{abstract}

\pacs{ 04.70.Dy, 95.30.Sf, 97.60.Lf } \maketitle

\newpage
\section{Introduction}

The Kaluza-Klein (KK) black holes with squashed horizons are a kind of
interesting Kaluza-Klein type metrics with the special topology and asymptotical structure \cite{IM,sq2,sq3,sq4,TVN,TW,sq6,SHKT}. This family of black
holes behave as fully five-dimensional black holes in the vicinity of horizon, while behave as four-dimensional black holes  with a constant twisted $S^1$ fiber in the far region. In these black holes, the size of compactified extra dimension can be adjustable by a parameter $r_{\infty}$. Recent investigations show that the spectrum of Hawking radiation \cite{sq1hw,sq2hw}, the quasinormal frequencies \cite{sqq1,sqq2} and the precession of a gyroscope in a circular orbit \cite{Ksq} in the background of the KK black holes with squashed horizons contain the information of the size of the extra dimension, which could open a possible window to observe extra dimensions in the future.

From the general theory of relativity, we know that photons would be deviated from their straight path when they pass close to a compact and massive body. The effects originating from the deflection of light rays in a gravitational field are known as gravitational lensing \cite{Einstein,schneider,Darwin}. Like a natural and large telescope, gravitational lensing can help us to capture the information about the very dim stars which are far away from our Galaxy. The
strong gravitational lensing is caused by a compact object with a photon sphere. As photons pass close to the photon sphere, the deflection angles of of the light rays become very large, which yields that there exist two infinite sets of faint relativistic images on each side of the black hole. With these relativistic images, we could extract the information about black holes in the Universe and verify profoundly alternative theories of gravity in their strong field regime \cite{Vir,Fritt,Bozza1,Eirc1,whisk,Bozza2,Bozza3,Bozza4,Gyulchev,Bhad1,TSa1,AnAv,gr1,gr2,gr3,Kraniotis}. Recently, we have studied the strong gravitational lensing in the background of a Schwarzschild squashed KK black hole \cite{schen} and a squashed KK G\"{o}del black hole \cite{schen2} and find that the size of the extra dimension and the
the global rotation parameter $j$ of the G\"{o}del Universe affects the photon sphere radius,  the deflection angle and the corresponding observational variables in the strong gravitational lensing.  Sadeghi \textit{et al} have considered the effect of the charge on the strong gravitational lensing in the squashed KK black holes \cite{JH,JAH}. These investigations
could help us to understand further the effects of the scale of extra dimension on the strong gravitational lensings. However, to my knowledge,
the strong gravitational lensing is still open in the background of a rotating squashed KK black hole. Since the rotation is a universal phenomenon for the celestial bodies in our Universe, it is necessary to study the strong gravitational lensing in the rotating squashed KK black hole spacetime. With the squashing transformation, Wang \cite{TW} obtained a rotating squashed KK black hole spacetime with two equal angular momenta in Einstein theory. This black hole solution with squashed horizons is geodesic complete and free of naked singularities. It has the similar
topology and asymptotical structure to that of the static squashed KK black hole, but with richer physical properties. In this paper, we are going to study the strong gravitational lensing in this rotating squashed KK black hole and probe the effects of the rotation parameter of black hole and
the scale of extra dimension  on the deflection angle and
the coefficients in the strong field limit.

The plan of our paper is organized as follows. In Sec.II we introduce briefly the rotating squashed KK black hole \cite{TW}. In Sec.III we adopt to Bozza's method \cite{Bozza2,Bozza3,Bozza4} and obtain the deflection angles for light rays propagating in the rotating squashed KK  black hole. In Sec.IV we suppose that the gravitational field of the supermassive black hole at the center of our Galaxy can be described by this metric and then obtain the numerical results for the main observables in the strong
gravitational lensing. Moreover, we also make a comparison among the strong gravitational lensings in the rotating squashed KK, the squashed KK G\"{o}del and four-dimensional Kerr black hole spacetimes. At last, we present a summary.

\section{The Rotating Kaluza-Klein Black Hole Spacetime With Squashed Horizon}
Let us now review briefly a rotating squashed KK black hole without charge, which can be obtained  by applying the squashing
transformation  techniques to a five-dimensional Kerr black hole with two equal angular momenta \cite{TW}. In terms of Meurer-Cartan 1-forms, the metric of a rotating KK black bole has a form
\begin{eqnarray}\label{metric0}
ds^2=-d\tilde{t}^2+\frac{\Sigma_0}{\Delta_0}k(r)^2dr^2+\frac{r^2+a^2}{4}
[k(r)(\sigma^2_1+\sigma^2_2)+\sigma^2_3]+\frac{M}{r^2+a^2}(d\tilde{t}-\frac{a}{2}\sigma_3)^2,
\end{eqnarray}
with
\begin{eqnarray}
\sigma_1&=&-\sin{\tilde{\psi}} d\theta+\cos{\tilde{\psi}} \sin{\theta}d\phi,\nonumber\\
\sigma_2&=&\cos{\tilde{\psi}} d\theta+\sin{\tilde{\psi}} \sin{\theta}d\phi,\nonumber\\
\sigma_3&=&d\tilde{\psi}+\cos{\theta}d\phi,
\end{eqnarray}
where $0<\theta<\pi$, $0<\phi<2\pi$ and $0<\tilde{\psi}<4\pi$. The
parameters are given by
\begin{eqnarray}
\Sigma_0&=&r^2(r^2+a^2),\nonumber\\
\Delta_0&=&(r^2-r^2_{+})(r^2-r^2_{-}),\nonumber\\
k(r)&=&\frac{(r^2_{\infty}-r^2_{+})(r^2_{\infty}-r^2_{-})}{(r^2_{\infty}-r^2)^2}.
\end{eqnarray}
The quantities $M$ and $a$ are related to the mass and angular momenta of black hole, respectively. $r_{\infty}$ corresponds to the spatial infinity. The polar coordinate $r$ is limited in the range $0<r<r_{\infty}$.
The outer and inner horizons are located at
$r=r_+$ and $r=r_-$, which are relate to the parameters $M $, $a$ by $a^4=(r_+r_-)^2$ and $ M-2a^2=r^2_++r^2_-$. The shape of black
hole horizon is deformed by the parameter $k(r_+)$.

In this black hole spacetime (\ref{metric0}), the intrinsic singularity is the just one at $r=0$, while $r_{\pm}$ and $r_{\infty}$ are coordinate singularities. As in \cite{TW}, one can introduce
 a new radial coordinate
 \begin{eqnarray}
\rho=\tilde{\rho}_0\frac{{r}^2}{{r}^2_{\infty}-{r}^2},\label{r4}
\end{eqnarray}
with
\begin{eqnarray}
\tilde{\rho}^2_0=\frac{(r^2_{\infty}+a^2)[(r^2_{\infty}+a^2)^2-Mr^2_{\infty}]}{4r^4_{\infty}}
,
\end{eqnarray}
and then rewrite the metric (\ref{metric0}) as
 \begin{eqnarray}
ds^2=-d\tilde{t}^2+Ud\rho^2+R^2(\sigma _1^2+\sigma _2^2)+W^2\sigma_3^2+V(d\tilde{t}-\frac{a}{2}\sigma_3)^2,
\label{metric1}
\end{eqnarray}
where
 \begin{eqnarray}
K^2&=&\frac{\rho+\tilde{\rho}_0}{\rho+\frac{a^2}{r^2_{\infty}+a^2}\tilde{\rho}_0},\;\;\;\;\;\;\;\;V=\frac{M}{r_\infty^2+a^2}K^2,\;\;\;\;\;\;\;
W^2=\frac{r^2_{\infty}+a^2}{4K^2},\nonumber\\R^2&=&\frac{(\rho+\tilde{\rho}_0)^2}{K^2},\;\;\;\;\;\;\;
U=\bigg(\frac{r^2_\infty}{r^2_{\infty}+a^2}\bigg)^2\frac{\tilde{\rho}_0^2}{W^2-\frac{r_\infty^2}{4}\frac{\rho}{\rho+\tilde{\rho}_0}V}.
\end{eqnarray}
The parameter $\tilde{\rho}_0$ is a scale of transition from five-dimensional
spacetime to an effective four-dimensional one.
As the rotation parameter $a$ tends to zero, one can find that the metric (\ref{metric1}) reduces to that of a five-dimensional Schwarzschild black hole with squashed horizon. In the limit $\rho\rightarrow \infty$, i.e, $r\rightarrow r_{\infty}$, it is easy to find that there is a cross-term between $d\tilde{t}$ and $\sigma_3$ in the asymptotic form of the metric (\ref{metric1}). However, this cross-term can be vanished by
changing the coordinates as
\cite{TW}
\begin{eqnarray}
\tilde{t}=h~ t,\;\;\;\;\;\tilde{\psi}=\psi-j ~t.
\end{eqnarray}
where
\begin{eqnarray}
h=\sqrt{\frac{(r_\infty^2+a^2)^2-Mr_\infty^2}{(r_\infty^2+a^2)^2+Ma^2}},\;\;\;\;\;\;\;
j=\frac{2Ma}{(r_\infty^2+a^2)^2+Ma^2}.
\label{newsys}
\end{eqnarray}
This means that the asymptotic topology of the spacetime (\ref{metric1}) is the same as that of the Schwarzschild squashed KK black hole spacetime.
The Komar mass $M_k$ of the rotating squashed KK black hole (\ref{metric1}) can be given by \cite{Komar,Komar1}
\begin{eqnarray}
M_k&=&\frac{M\pi}{2 G_5}\frac{\left(r_{\infty}^2+a^2\right)^2-M a^2}{ \sqrt{\left(r_{\infty}^2+a^2\right)^2+M a^2}\sqrt{\left(r_{\infty}^2+a^2\right)^2-Mr_{\infty}^2}}\nonumber\\&=&
\frac{M}{4 G_4}\frac{\left(r_{\infty}^2+a^2\right)^2-M a^2}{ \left(r_{\infty}^2+a^2\right)^2+M a^2}\frac{\sqrt{r_{\infty}^2+a^2}}{\sqrt{\left(r_{\infty}^2+a^2\right)^2-Mr_{\infty}^2}},
\label{Mk}
\end{eqnarray}
where $G_5$ and $G_4$ are the five-dimensional and four-dimensional gravitational constants, respectively. Therefore, in the rotating squashed KK black hole spacetime, the relationship between $G_5$ and $G_4$ can be expressed as
\begin{eqnarray}
G_5=2\pi r'_{\infty}G_4,\label{NG}
\end{eqnarray}
with
\begin{eqnarray}
r'_{\infty}=\sqrt{\frac{(r^2_\infty+a^2)^2+Ma^2}{r_\infty^2+a^2}}.
\end{eqnarray}
The expression of $r'_{\infty}$ is more complicated than that of $r_{\infty}$. However, in the rotating squashed KK black hole spacetime (\ref{metric1}), one can find \cite{TW} that the parameter $r'_{\infty}$ for the compactified dimension is better than $r_{\infty}$ because the geometric interpretation is clearer for $r'_{\infty}$
than for $r_{\infty}$. As $a$ disappears, we find that $r'_{\infty}$ reduces to $r_{\infty}$ and then
the relationship (\ref{NG}) tend to the usual form ( i.e., $G_5=2\pi r_{\infty}G_4$ ) in the Schwarzschild squashed KK black hole spacetime. As in Ref.\cite{Ksq}, we can rewritten Komar mass as $M_k=\frac{\pi r'_{\infty} \rho_M}{G_5}=\frac{\rho_M}{2G_4}$, which implies that $\rho_M$ can be expressed as
\begin{eqnarray}
\rho_M=2G_4M_k=
\frac{M}{2}\frac{\left(r_{\infty}^2+a^2\right)^2-M a^2}{ \left(r_{\infty}^2+a^2\right)^2+M a^2}\frac{\sqrt{r_{\infty}^2+a^2}}{
\sqrt{\left(r_{\infty}^2+a^2\right)^2-Mr_{\infty}^2}}.\label{rho1}
\end{eqnarray}
In order to simplify the calculation, we here introduce a new radial coordinate change
\begin{eqnarray}
{r'}^2=\frac{(r^2_\infty+a^2)^2+Ma^2}{r_\infty^4+r^2a^2}r^2,
\label{nr}
\end{eqnarray}
a new scale of extra dimension
\begin{eqnarray}
\rho^2_0=\frac{{r'}^2_{\infty}-M}{4},
\label{rho0}
\end{eqnarray}
and a new rotation parameter
\begin{eqnarray}
b=\sqrt{\frac{M}{r_{\infty}^2+a^2}}\;a,
\label{bb}
\end{eqnarray}
we find that the radial coordinate (\ref{r4}) and the quantity $\rho_M$ can be rewritten as
\begin{eqnarray}
\rho=\rho_0\frac{{r'}^2}{{r'}^2_{\infty}-{r'}^2},
\end{eqnarray}
and
\begin{eqnarray}
\rho_M=\frac{\rho_0 M}{r'^2_{\infty}-M}\frac{\left(r_{\infty}^2+a^2\right)^2-M a^2}{ \left(r_{\infty}^2+a^2\right)^2+M a^2}=\frac{M\rho_0}{{r'}_\infty^2-M}(1-\frac{2b^2}{{r'}_\infty^2}),
\end{eqnarray}
respectively.
With help of these operation, one can find that the metric of the rotating squashed KK black hole spacetime (\ref{metric1}) can be expressed as
\begin{eqnarray}
ds^2=-A(\rho)dt^2+B(\rho)d\rho^2+C(\rho)(d\theta^2+\sin^2\theta d\phi^2)+D(\rho)(d\psi+\cos\theta d\phi)^2-2H(\rho)dt(d\psi+\cos\theta d\phi),
\label{metricn1}
\end{eqnarray}
with
\begin{eqnarray}
A(\rho)&=&\frac{4-\mathcal{V}(a j-2)^2-4j^2\mathcal{W}^2}{4h^2},\;\;\;\;\;\;\;B(\rho)=\mathcal{U}(\rho),\;\;\;\;\;\;\;\;\;\;\;
C(\rho)=\mathcal{R}^2(\rho),\nonumber\\ D(\rho)&=&\frac{a^2\mathcal{V}+4\mathcal{W}^2}{4},\;\;\;\;\;\;
H(\rho)=\frac{2a\mathcal{V}-(a^2\mathcal{V}+4\mathcal{W}^2)j}{4h},
\end{eqnarray}
where
\begin{eqnarray}
\mathcal{K}^2&=&\frac{\rho+\frac{r^2_{\infty}+a^2}{r^2_{\infty}}\rho_0}{\rho
+\frac{a^2}{r^2_{\infty}}\rho_0},\;\;\;\;\;\;\;\;
\mathcal{V}=\frac{M}{r_\infty^2+a^2}\mathcal{K}^2,\;\;\;\;\;\;\;
\mathcal{W}^2=\frac{r^2_{\infty}+a^2}{4\mathcal{K}^2},\nonumber\\
\mathcal{R}^2(\rho)&=&(\rho+\frac{a^2}{r^2_{\infty}}\rho_0)
(\rho+\frac{r^2_{\infty}+a^2}{r^2_{\infty}}\rho_0),\;\;\;\;\;\;\;
\mathcal{U}(\rho)=\frac{\rho_0^2}{\mathcal{W}^2-\frac{r_\infty^2}{4}
\frac{\rho}{\mathcal{K}\mathcal{R}}\mathcal{V}}.
\end{eqnarray}
Among the parameters $\rho_0$,  $M$,  $a$,  $r_{\infty}$,  $r'_{\infty}$, $\rho_M$ and $b$, there are only three independent parameters. For simplicity, we chose the parameters $\rho_0$, $\rho_M$ and $b$ as the independent parameters. The others are related to the selected three parameters by
\begin{eqnarray}
r_\infty^2&=&2\rho_0(\rho_M+\rho_0)-b^2+\sqrt{b^4-4b^2\rho_0\left(\rho_0-\rho_M\right)+
4\rho_0^2\left(\rho_M+\rho_0\right)^2}
\nonumber\\&&-\frac{b^2(4\rho_0^2-b^2)}{2\rho_0(\rho_M-\rho_0)+b^2+\sqrt{b^4-4b^2\rho_0\left(\rho_0-\rho_M\right)
+4\rho_0^2\left(\rho_M+\rho_0\right)^2}},\nonumber\\
{r'}_\infty^2&=&b^2+2\rho_0\rho_M+2\rho_0^2+\sqrt{b^4-4b^2\rho_0
\left(\rho_0-\rho_M\right)+4\rho_0^2\left(\rho_M+\rho_0\right)^2},\nonumber\\
M&=&b^2+2\rho_0\rho_M-2\rho_0^2+\sqrt{b^4-4b^2\rho_0\left(\rho_0-
\rho_M\right)+4\rho_0^2\left(\rho_M+\rho_0\right)^2},\nonumber\\
a&=&b\bigg[1+\frac{4\rho_0^2-b^2}{2\rho_0(\rho_M-\rho_0)+b^2+\sqrt{b^4-4b^2\rho_0\left(\rho_0-\rho_M\right)
+4\rho_0^2\left(\rho_M+\rho_0\right)^2}}\bigg]^{\frac{1}{2}}.
\end{eqnarray}
With these quantities, we can find that all of coefficients in the metric (\ref{metricn1}) can be expressed as the functions of the parameters $\rho_0$, $\rho_M$ and $b$, which means that we can  study the strong gravitational lensing in the rotating squashed KK black hole spacetime (\ref{metric1}) through the standard form used in \cite{Vir,Fritt,Bozza1,Eirc1,whisk,Bozza2,Bozza3,Bozza4,Gyulchev,Bhad1,TSa1,AnAv,gr1,gr2,gr3,Kraniotis}.

\section{Deflection angle in a rotating squashed Kaluza-Klein black hole spacetime}
In this section, we will study deflection angles of the light rays
when they pass close to a rotating squashed KK black hole, and then probe the effects of the rotation parameter $b$ and
the scale of extra dimension $\rho_0$ on the deflection angle and
the coefficients in the strong field limit. For simplicity, we
here just consider that both the observer and the source lie in the equatorial plane in the rotating squashed KK black hole spacetime (\ref{metricn1}) and the whole trajectory of the photon is limited on the same plane
With this condition $\theta=\frac{\pi}{2}$, we get the reduced
metric in the form
\begin{eqnarray}
ds^2=-A(\rho)dt^2+B(\rho)d\rho^2+C(\rho)d\phi^2
+D(\rho)d\psi^2-2H(\rho)dtd\psi.\label{l1}
\end{eqnarray}
From the null geodesics, it is easy to obtain three constants of motion
\begin{eqnarray}
E&=-&g_{0\mu}\dot{x}^{\mu}=A(\rho)\dot{t}+H(\rho)\dot{\psi},\nonumber\\
L_{\phi}&=&g_{3\mu}\dot{x}^{\mu}=C(\rho)\dot{\phi},\nonumber\\
L_{\psi}&=&g_{4\mu}\dot{x}^{\mu}=D(\rho)\dot{\psi}-H(\rho)\dot{t}.
\end{eqnarray}
where a dot represents a derivative with respect to affine parameter
$\lambda$ along the geodesics. $E$ is the energy of the phone, $L_{\phi}$ and $L_{\psi}$ correspond to its angular momentum in the $\phi$ and $\psi$ directions, respectively. Making use of these three constants, one can
find that the equations of motion of the photon can be expressed further as
\begin{eqnarray}
\frac{dt}{d\lambda}&=&\frac{D(\rho)E-H(\rho)L_{\psi}}{H^2(\rho)+A(\rho)D(\rho)},\nonumber\\
\frac{d\phi}{d\lambda}&=&\frac{L_{\phi}}{C(\rho)},\nonumber\\
\frac{d\psi}{d\lambda}&=&\frac{H(\rho)E+A(\rho)L_{\psi}}{H^2(\rho)+A(\rho)D(\rho)}.
\end{eqnarray}
\begin{eqnarray}
\bigg(\frac{d\rho}{d\lambda}\bigg)^2
=\frac{1}{B(\rho)}\bigg[\frac{D(\rho)E-2H(\rho)EL_{\psi}-A(\rho)L^2_{\psi}}{H^2(\rho)+A(\rho)D(\rho)}-\frac{L^2_{\phi}}{C(\rho)}
\bigg].
\end{eqnarray}
Considering the $\theta$-component of the null geodesics in the equatorial
plane ($\theta=\frac{\pi}{2}$), we have
\begin{eqnarray}
\frac{d\phi}{d\lambda}\bigg[D(\rho)\frac{d\psi}{d\lambda}-H(\rho)\frac{dt}{d\lambda}\bigg]=0,
\end{eqnarray}
which implies that either $\frac{d\phi}{d\lambda}=0$ or
$L_{\psi}=D(\rho)\frac{d\psi}{d\lambda}-H(\rho)\frac{dt}{d\lambda}=0$.
As done in the usual squashed KK  black hole spacetimes
\cite{schen,schen2}, here we set $L_{\psi}=0$, which means that the total
angular momentum $J$ of the photo is equal to the constant
$L_{\phi}$ and the effective potential for the
photon passing close to the black hole can be written as
\begin{eqnarray}
V(\rho)=\frac{1}{B(\rho)}\bigg[\frac{D(\rho)E}{H^2(\rho)+A(\rho)D(\rho)}-\frac{L^2_{\phi}}{C(\rho)}
\bigg].
\end{eqnarray}
With this effective potential, one can obtain that the
impact parameter and the equation of circular photon orbits are
\begin{eqnarray}
u=J=\sqrt{\frac{C(\rho)D(\rho)}{H(\rho)^2+A(\rho)D(\rho)}},\label{u1}
\end{eqnarray}
and
\begin{eqnarray}
D(\rho)\bigg[H(\rho)^2+A(\rho)D(\rho)\bigg]C'(\rho)-C(\rho)\bigg[D(\rho)^2A'(\rho)
+2D(\rho)H(\rho)H'(\rho)-H(\rho)^2D'(\rho)\bigg]=0,\label{phs-e}
\end{eqnarray}
respectively. Here we set $E=1$. The equations (\ref{u1}) and
(\ref{phs-e}) are similar to those in the squashed KK G\"{odel} black hole spacetime because their metric have similar forms in the equatorial plane, but they are more complex than those in the usual spherical
symmetric black hole spacetime.  As the rotation parameter
$b\rightarrow 0$, we find that the function $H(\rho)\rightarrow 0$,
which yields that the impact parameter (\ref{u1}) and the equation of circular photon orbits (\ref{phs-e}) reduce to those in the usual Schwarzschild
squashed KK black hole spacetime \cite{schen}. The biggest real root external to the horizon of equation (\ref{phs-e}) defines the marginally stable circular radius of photon. For the rotating squashed KK black hole spacetime (\ref{metricn1}), the equation of circular photon orbits takes the
form
\begin{eqnarray}
\mathcal{A}{\rho'}^5+\mathcal{B}{\rho'}^4+\mathcal{C}{\rho'}^3+\mathcal{D}{\rho'}^2+\mathcal{E}{\rho'}+\mathcal{F}=0
\label{phsp}
\end{eqnarray}
where the variable $\rho'$ is related to $\rho$ by
\begin{eqnarray}
\rho'&=&\frac{\rho}{\rho_0}+\frac{b^2}{2\rho_0\rho_M-2\rho_0^2+\sqrt{b^4-4b^2\rho_0\left(\rho_0-
\rho_M\right)+4\rho_0^2\left(\rho_M+\rho_0\right)^2}},
\end{eqnarray}
and the coefficients are
\begin{eqnarray}
\mathcal{A}&=&8\rho_0^2\left(b^2+2\rho_0\rho_M+2\rho_0^2
+\sqrt{b^4-4b^2\rho_0\left(\rho_0-
\rho_M\right)+4\rho_0^2\left(\rho_M+\rho_0\right)^2}\right),\nonumber\\
\mathcal{B}&=&\left(6b^2-6\rho_0\rho_M+10\rho_0^2-3\sqrt{b^4-4b^2\rho_0\left(\rho_0-
\rho_M\right)+4\rho_0^2\left(\rho_M+\rho_0\right)^2}\right)\times\nonumber\\
&&\left(3b^2+2\rho_0\rho_M+2\rho_0^2+\sqrt{b^4-4b^2\rho_0\left(\rho_0-
\rho_M\right)+4\rho_0^2\left(\rho_M+\rho_0\right)^2}\right)-12b^4,\nonumber\\
\mathcal{C}&=&2\left(3b^2-2\rho_0\rho_M+2\rho_0^2-\sqrt{b^4-4b^2\rho_0\left(\rho_0-
\rho_M\right)+4\rho_0^2\left(\rho_M+\rho_0\right)^2}\right)\times\nonumber\\
&&\left(3b^2+2\rho_0\rho_M+2\rho_0^2+\sqrt{b^4-4b^2\rho_0\left(\rho_0-
\rho_M\right)+4\rho_0^2\left(\rho_M+\rho_0\right)^2}\right)-8b^4,\nonumber\\
\mathcal{D}&=&4b^2\left(3b^2+2\rho_0^2\right),\;\;\;\;\;\;\;\;\;\;
\mathcal{E}=6b^4,\;\;\;\;\; \mathcal{F}=b^4.
\label{phspco}
\end{eqnarray}
Obviously, this equation depends on both the rotation parameter $b$ and the scale of transition $\rho_0$. The presence of the rotation parameter $b$ makes the equation more complex so that it is
impossible to get an analytical form for the marginally circular photon orbit radius in this case. As $b\rightarrow 0$, we can find that since the coefficients $\mathcal{D}$, $\mathcal{E}$ and $\mathcal{F}$ vanish the Eq.(\ref{phsp}) reduces to a quadratic equation
\begin{eqnarray}
2\rho^2+(\rho_0-3\rho_M)\rho-2\rho_0\rho_M=0,
\end{eqnarray}
and the marginally circular photon orbit radius becomes $\rho_{ps}=\frac{3\rho_M-\rho_0+\sqrt{9\rho^2_M+10\rho_M\rho_0+\rho^2_0}}{4}$, which is consistent with that in the Schwarzschild squashed  KK black hole \cite{schen}. As $\rho_0\rightarrow 0$, one can get $\rho_{ps}=\frac{1}{4}(3\rho_M+\sqrt{9\rho^2_M-8b^2} )$, which decreases with the rotation parameter $b$ and tends to $\frac{3}{2}\rho_M$ as $b$ disappears.  Moreover, in the limit $\rho_0\rightarrow\infty$, we have
\begin{eqnarray}
\rho_{ps}&=&\frac{2\rho_M-b^2}{4\rho_M}+\frac{1}{2}
\sqrt{\frac{\mathcal{Q}^2+(6\rho^2_M-2b^2)\mathcal{Q}+4b^4}{6\mathcal{Q}}}\nonumber\\&+&\frac{1}{2}\bigg[
\frac{\mathcal{Q}(12\rho^2_M -\mathcal{Q})
-4b^2(\mathcal{Q}+b^2)}{6\mathcal{Q}}+\frac{\sqrt{6\mathcal{Q}}
\rho_M(2\rho_M-b^2)}{\sqrt{
\mathcal{Q}^2+(6\rho^2_M-2b^2)\mathcal{Q}+4b^4}}\bigg]^{\frac{1}{2}},
\end{eqnarray}
with
\begin{eqnarray}
\mathcal{Q}=b^{\frac{4}{3}}\bigg[27\rho^2_M-8b^2+3\rho_M\sqrt{81\rho^2_M-48b^2}
\bigg]^{\frac{1}{3}}.
\end{eqnarray}
Obviously, it also decreases with the rotation parameter $b$.
In Fig.(1), we set $\rho_M=1$ and plot the variety of the marginally stable circular radius of photon $\rho_{ps}$ with the parameters $b$ and $\rho_0$. It shows that
with increase of the scale of transition $\rho_0$, $\rho_{ps} $ first
decreases and then increases in the  rotating squashed KK black hole background, but increases monotonously in the Schwarzschild squashed KK black hole spacetime ( i.e., $b=0$ ). In the squashed KK G\"{o}del black hole spacetime, we find \cite{schen2} that with the increase of $\rho_0$, $\rho_{ps} $ increases for the smaller global rotation parameter  $j$ and decreases for the larger one, which implies that the effects of  the rotation parameter $b$ of the black hole itself on the gravitational lensing
is different from that of the global rotation parameter $j$ of
the G\"{o}del Universe background. For fixed $\rho_0$, it is easy to obtain that $\rho_{ps}$ decreases monotonically with
the increase of the rotation parameter $b$. Moreover, we know that in the Kerr black hole, the marginally stable circular radius of photon are different for the photons winding in the same direction (i.e., $a>0$) or in the converse direction
(i.e., $a <0$) of the black hole rotation. However, one can obtain that in the rotating squashed KK black hole spacetime the marginally stable circular radius of photon $\rho_{ps}$ is independent of the sign of the rotation parameter $b$
since all of quantities in Eqs.(\ref{phsp}) and (\ref{phspco}) are the function of $b^2$. Thus, the marginally stable circular radius of photon
in the rotating squashed KK black hole spacetime keeps the same whether the photon moves in the same or converse direction of
the  rotation of the black hole. It is similar to that in the squashed KK G\"{o}del black hole spacetime, but it is different from that in the usual Kerr black hole spacetime.
\begin{figure}[ht]
\begin{center}
\includegraphics[width=6cm]{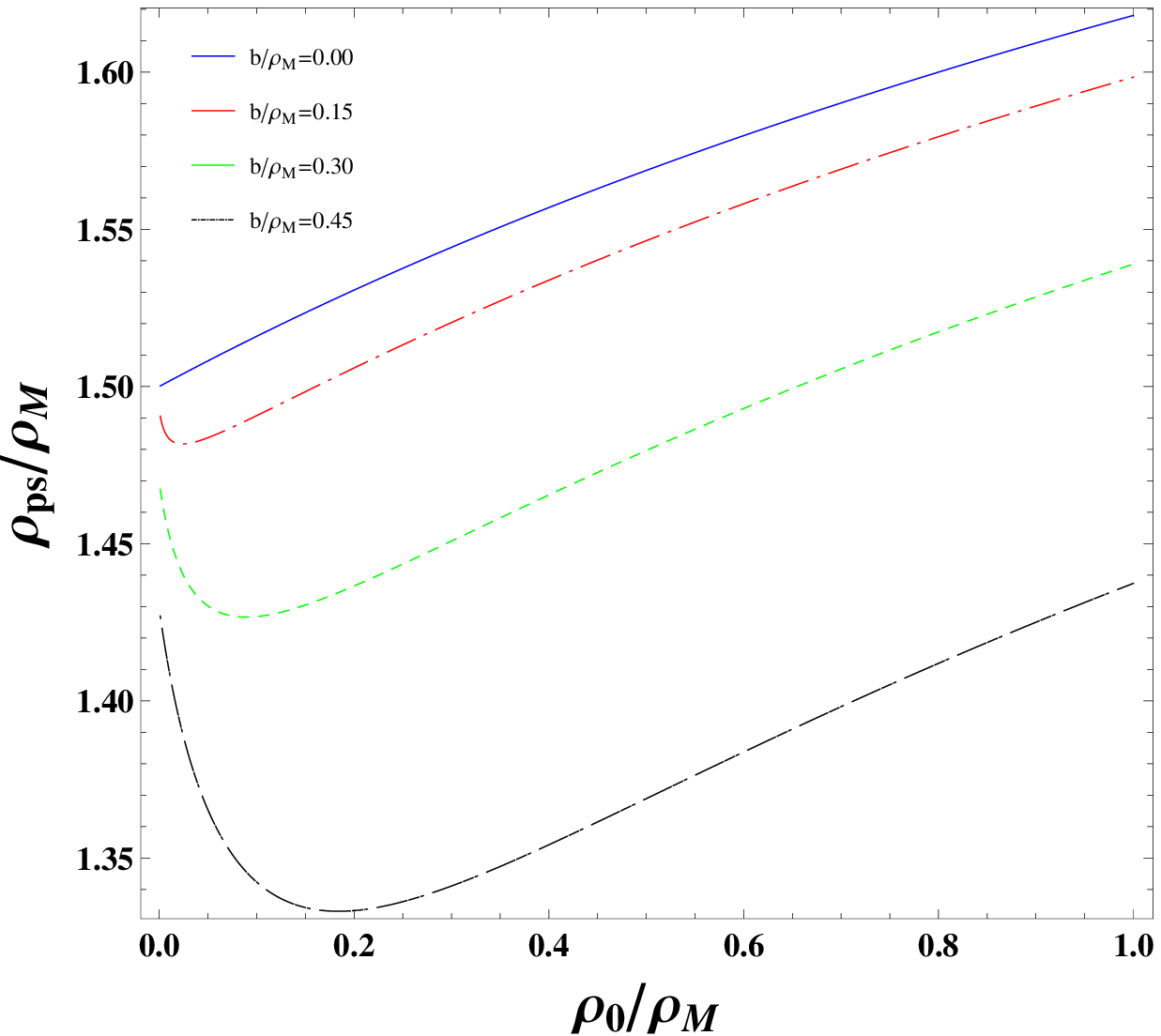}\;\;\;\includegraphics[width=6cm]{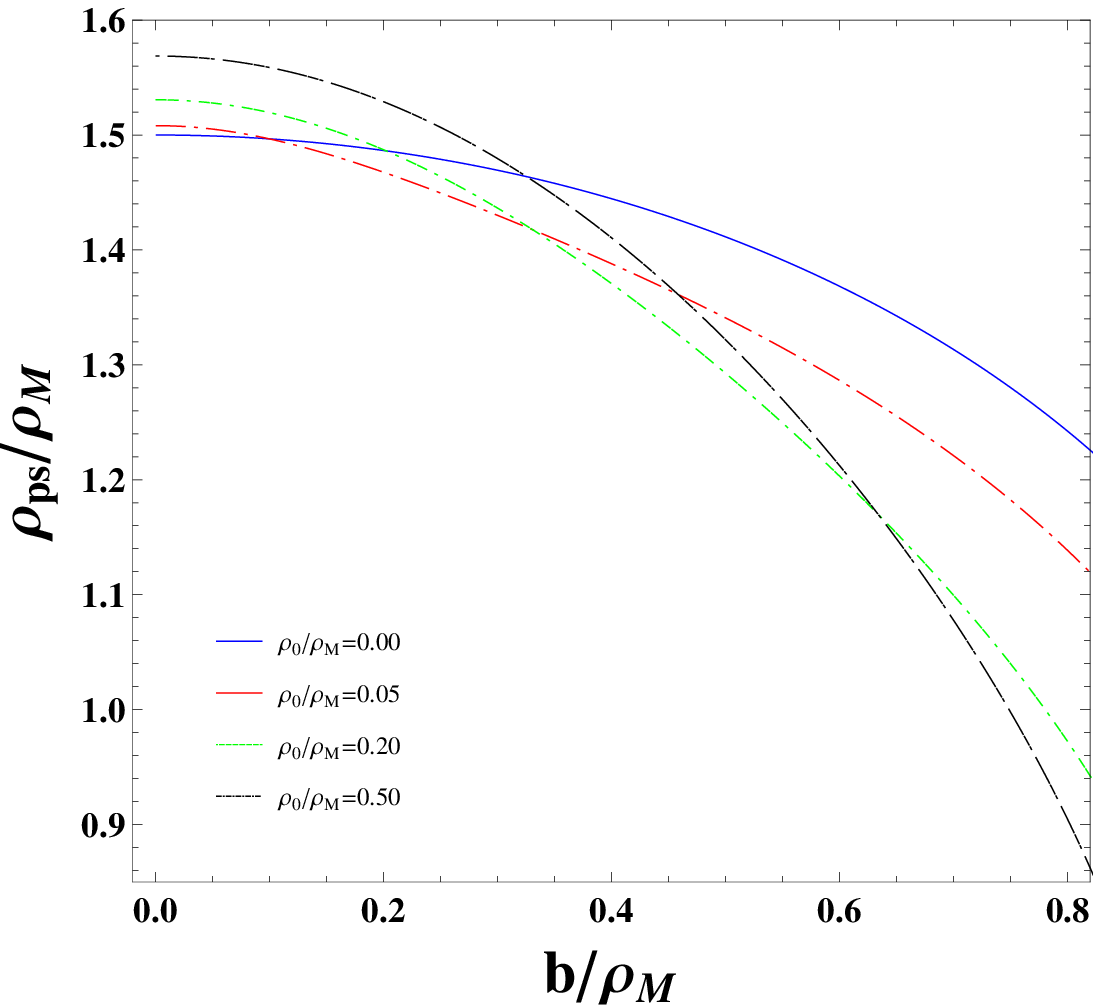}
\caption{Variety of the marginally stable circular radius of photon $\rho_{ps}$ with the parameters $\rho_0/\rho_M$ and $b/\rho_M$ in a rotating squashed KK black hole spacetime.}
\end{center}
\end{figure}

The deflection angles $\phi$ and $\psi$  for the photon coming from infinite in a rotating KK black hole spacetime can be expressed as
\begin{eqnarray}
\alpha_{\phi}(\rho_{s})&=&I_{\phi}(\rho_s)-\pi,\label{dphi}\nonumber\\
\alpha_{\psi}(\rho_{s})&=&I_{\psi}(\rho_s)-\pi,\label{dpsi}
\end{eqnarray}
respectively.  The quantities $I_{\phi}(\rho_s)$ and $I_{\psi}(\rho_s)$ have the forms
\begin{eqnarray}
I_{\phi}(\rho_s) &=&2\int^{\infty}_{\rho_s}\frac{\sqrt{B(\rho)
\mathcal{F}(\rho)C(\rho_s)}}{C(\rho)}\frac{1}{\sqrt{\mathcal{F}(\rho_s)-
\frac{\mathcal{F}(\rho)C(\rho_s)}{C(\rho)}}}d\rho,\label{int1}\\
I_{\psi}(\rho_s)&=&2\int^{\infty}_{\rho_s}\frac{H(\rho)}{D(\rho)}\sqrt{\frac{B(\rho)\mathcal{F}(\rho_s)}{\mathcal{F}(\rho)}}\frac{1}{\sqrt{\mathcal{F}(\rho_s)-
\frac{\mathcal{F}(\rho)C(\rho_s)}{C(\rho)}}}d\rho,\label{int2}
\end{eqnarray}
with
\begin{eqnarray}
\mathcal{F}(\rho)=\frac{H^2(\rho)+A(\rho)D(\rho)}{D(\rho)},
\end{eqnarray}
where $\rho_s$ is the closest approach distance of the light ray.
It is clear that both of the deflection angles
increase when parameter $\rho_s$ decreases. If $\rho_s$ is equal to
the marginally stable circular radius of photon $\rho_{ps}$, one can find that both
of the deflection angles becomes unboundedly large and the photon is captured in a circular orbit around the black hole. Let us now discuss the behavior of the deflection angles of the light rays in a rotating squashed KK black hole spacetime. It is
interesting to note that the deflection angle $\alpha_{\phi}(\rho_{s})$ is independent of whether the photon goes
with or against the rotation of the black hole because
the integral $I_{\phi}(\rho_s)$  is
function of the rotation parameter $b^2$.
However, from Eq. (\ref{int2}), we find that the integral
$I_{\psi}(\rho_s)$ contains the factor $b$, which means that the deflection
angle $\alpha_{\psi}(\rho_{s})$ for the photon traveling in the same
direction as the rotation of the black hole is
different from that traveling in converse direction. This tells us that although the black hole has the rotation paraters both in the $\psi$ and  $\phi$ directions, in the equatorial plane
$\theta=\frac{\pi}{2}$ the rotation of the black hole is really in
the $\psi$ direction rather than in the $\phi$ direction, which is also shown in the induce metric (\ref{l1}) where the only cross-term is $dtd\psi$. It means that the gravitational lensing by the rotating squashed KK black hole is different from that of usual four-dimensional Kerr black hole, which could in theory help us to detect the extra dimension through the gravitational lens. When the rotation parameter $b$
vanishes, one can find that the function $H(\rho)=0$ and then the deflection angle of $\psi$ tends to zero, which reduces to that of in the usual Schwarzschild
squashed KK black hole spacetime \cite{schen}.

As in \cite{schen2}, we will focus only on investigating the deflection angle
in the $\phi$ direction when the light rays pass close to the
black hole in the equatorial plane since it could be observed really by
our astronomical experiments. On the other hand, it is very convenient for us to
compare with the results obtained in the usual four-dimensional black hole spacetimes. As in \cite{Bozza2,Bozza3}, one can define a variable $z=1-\frac{\rho_s}{\rho}$, and rewrite
Eq.(\ref{int1}) as
\begin{eqnarray}
I_{\phi}(\rho_s)&=&\int^{1}_{0}R(z,\rho_s)f(z,\rho_s)dz,\label{in1}
\end{eqnarray}
with
\begin{eqnarray}
R(z,\rho_s)&=&2\frac{\rho^2}{\rho_sC(\rho)}\sqrt{B(\rho)
\mathcal{F}(\rho)C(\rho_s)}\nonumber\\&=&4\rho_s r^2_{\infty}
\sqrt{\frac{(r^2_{\infty}+a^2)^2+Ma^2}{(r^2_{\infty}+a^2)^2-Mr^2_{\infty}}}
\sqrt{\frac{\rho_0(r^2_{\infty}+a^2)(\rho_sr^2_{\infty}+a^2\rho_0)
[\rho_sr^2_{\infty}+(r^2_{\infty}+a^2)\rho_0]}{[\rho_sr^2_{\infty}+(1-z)a^2\rho_0]
[\rho_sr^2_{\infty}+(1-z)(r^2_{\infty}+a^2)\rho_0]}}
\nonumber\\&\times&
\bigg\{[(r^2_{\infty}+a^2)^2+Ma^2]\rho_sr^4_{\infty}+2a^2r^2_{\infty}\rho_0\rho_s
(1-z)(r^2_{\infty}+a^2)(r^2_{\infty}+M+a^2)\nonumber\\&+&a^2\rho^2_0(r^2_{\infty}+a^2)(M+a^2)
(1-z)^2\bigg\}^{-1/2},
\end{eqnarray}
and
\begin{eqnarray}
f(z,\rho_s)&=&\frac{1}{\sqrt{\mathcal{F}(\rho_s)-\mathcal{F}(\rho)C(\rho_s)/C(\rho)}}.\label{ffq}
\end{eqnarray}
The function $R(z, \rho_s)$ is regular for all values of $z$ and $\rho_s$. From Eq.(\ref{ffq}), we find that the function $f(z, \rho_s)$
diverges as $z$ tends to zero, i.e., as the photon approaches the marginally circular photon orbit. Therefore, we can split the integral (\ref{in1})
into the divergent part $I_D(\rho_s)$ and the regular one
$I_R(\rho_s)$
\begin{eqnarray}
I_D(\rho_s)&=&\int^{1}_{0}R(0,\rho_{ps})f_0(z,\rho_s)dz, \nonumber\\
I_R(\rho_s)&=&\int^{1}_{0}[R(z,\rho_s)f(z,\rho_s)-R(0,\rho_{ps})f_0(z,\rho_s)]dz
\label{intbr}.
\end{eqnarray}
Following in refs.\cite{Bozza2,Bozza3}, we can expand the argument of the square root in $f(z,\rho_{s})$ to the
second order in $z$
\begin{eqnarray}
f_s(z,\rho_{s})=\frac{1}{\sqrt{p(\rho_{s})z+q(\rho_{s})z^2}},
\end{eqnarray}
with
\begin{eqnarray}
p(\rho_{s})&=&\frac{\rho_s}{C(\rho_s)}\bigg[C'(\rho_s)\mathcal{F}(\rho_s)-C(\rho_s)\mathcal{F}'(\rho_s)\bigg],  \nonumber\\
q(\rho_{s})&=&\frac{\rho^2_s}{2C(\rho_s)}\bigg[2C'(\rho_s)C(\rho_s)\mathcal{F}'(\rho_s)-2C'(\rho_s)^2\mathcal{F}(\rho_s)
+\mathcal{F}(\rho_s)C(\rho_s)C''(\rho_s)-C^2(\rho_s)\mathcal{F}''(\rho_s)\bigg].\label{pq}
\end{eqnarray}
Comparing Eq.(\ref{phs-e}) with Eq.(\ref{pq}), one can find that if $\rho_{s}$ tends to $\rho_{ps}$ the coefficient $p(\rho_s)$ vanishes. This means that the leading term of the divergence in $f_s(z,\rho_{s})$ is $z^{-1}$ and
the integral (\ref{in1}) diverges logarithmically. Thus, in the strong
field region, the deflection angle in the $\phi$ direction can be approximated very well as \cite{Bozza2}
\begin{eqnarray}
\alpha(\theta)=-\bar{a}\log{\bigg(\frac{\theta
D_{OL}}{u_{ps}}-1\bigg)}+\bar{b}+O(u-u_{ps}), \label{alf1}
\end{eqnarray}
\begin{figure}[ht]
\begin{center}
\includegraphics[width=6cm]{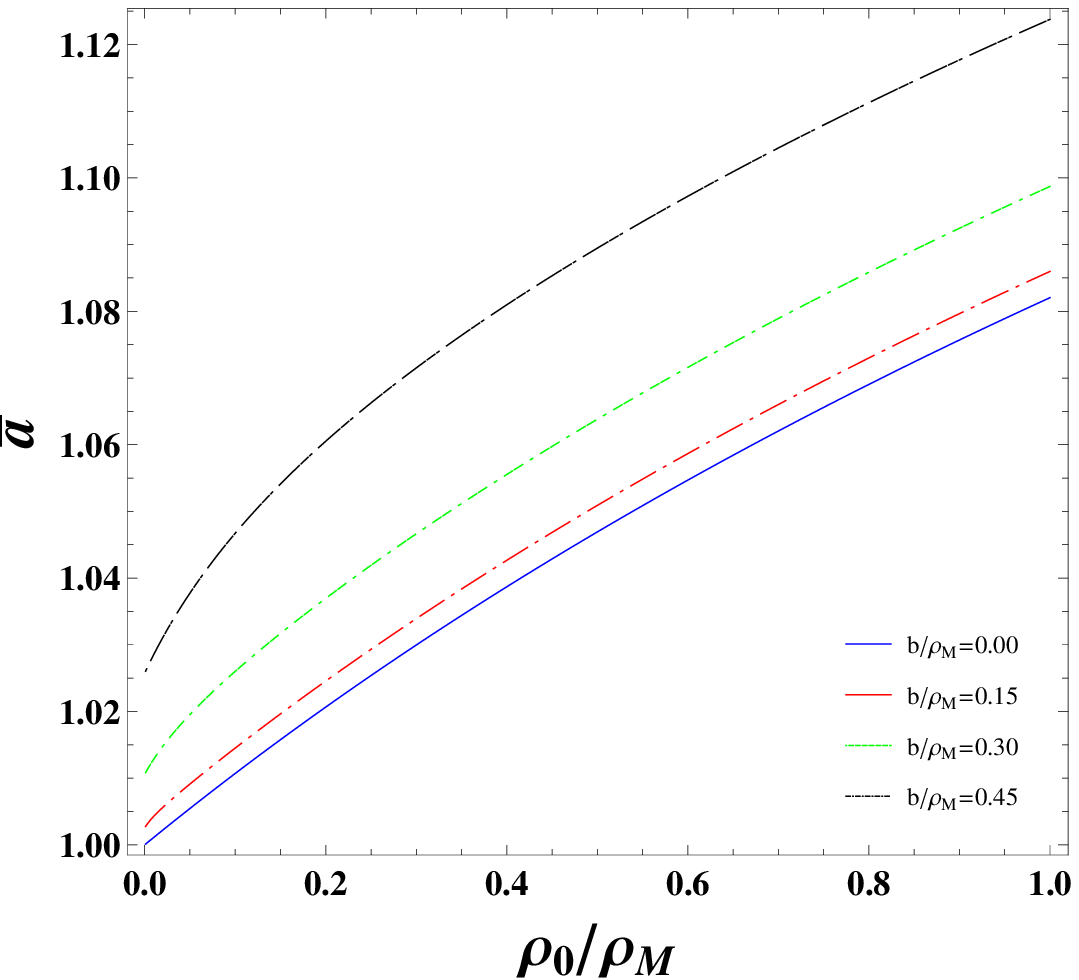}\;\;\;\includegraphics[width=6cm]{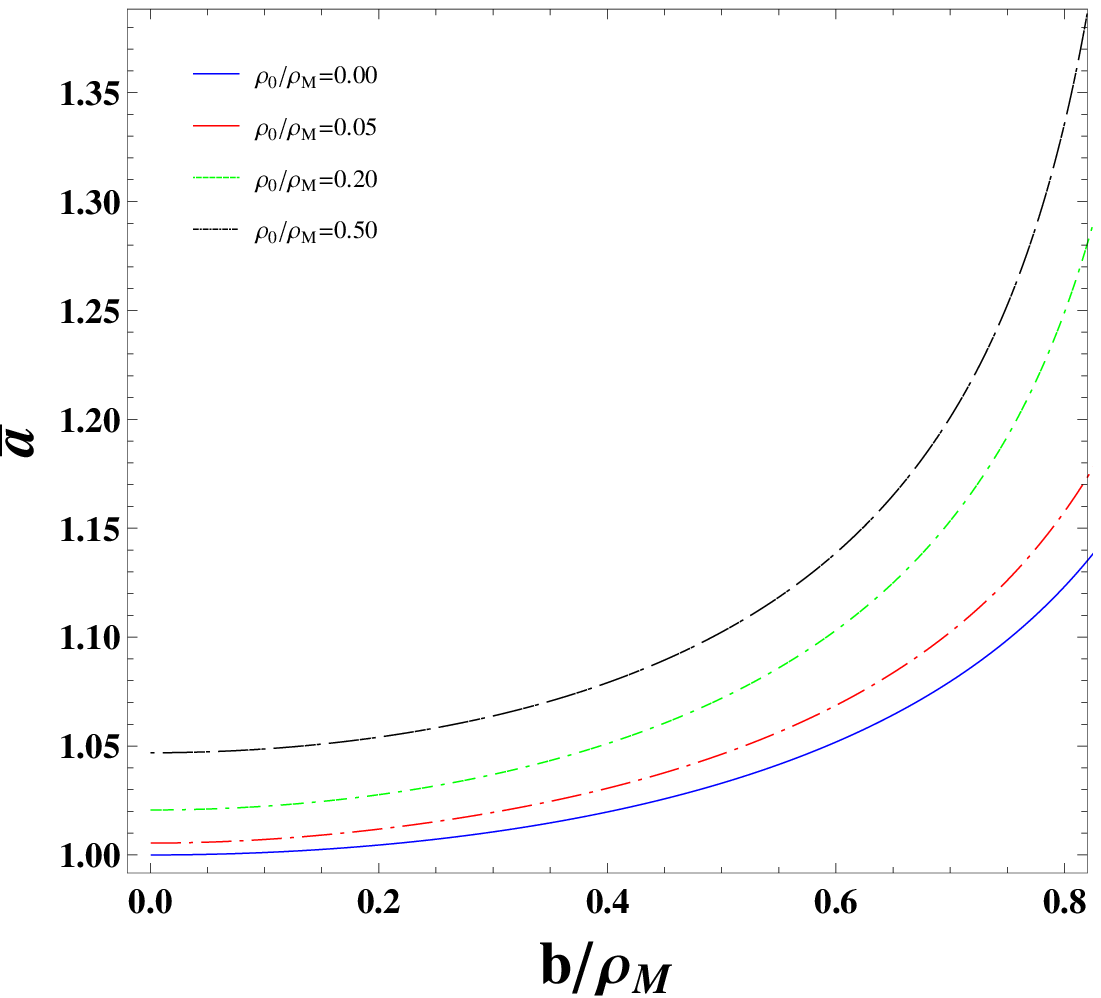}
\caption{Variety of the coefficient $\bar{a}$ with the parameters $\rho_0/\rho_M$ and $b/\rho_M$ in a rotating squashed KK black hole spacetime.}
\end{center}
\end{figure}
\begin{figure}[ht]
\begin{center}
\includegraphics[width=6cm]{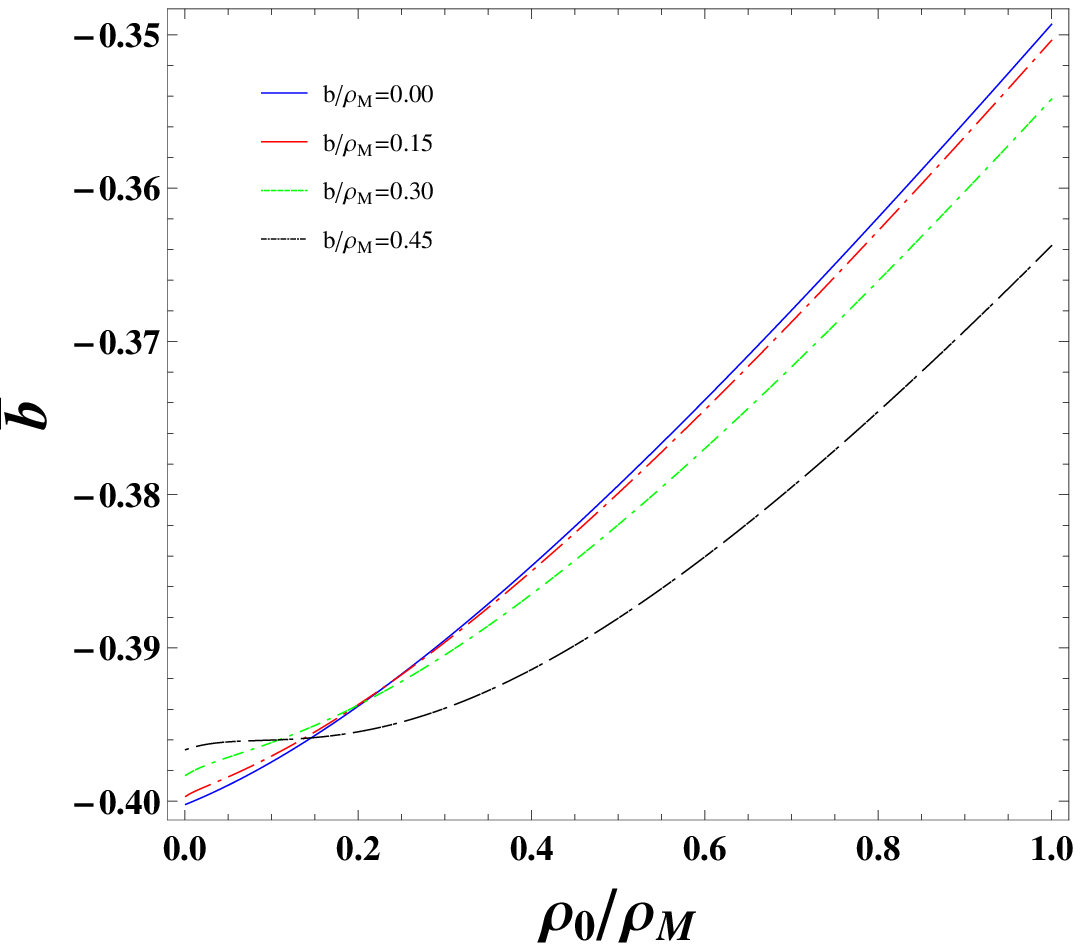}\;\;\;\includegraphics[width=6cm]{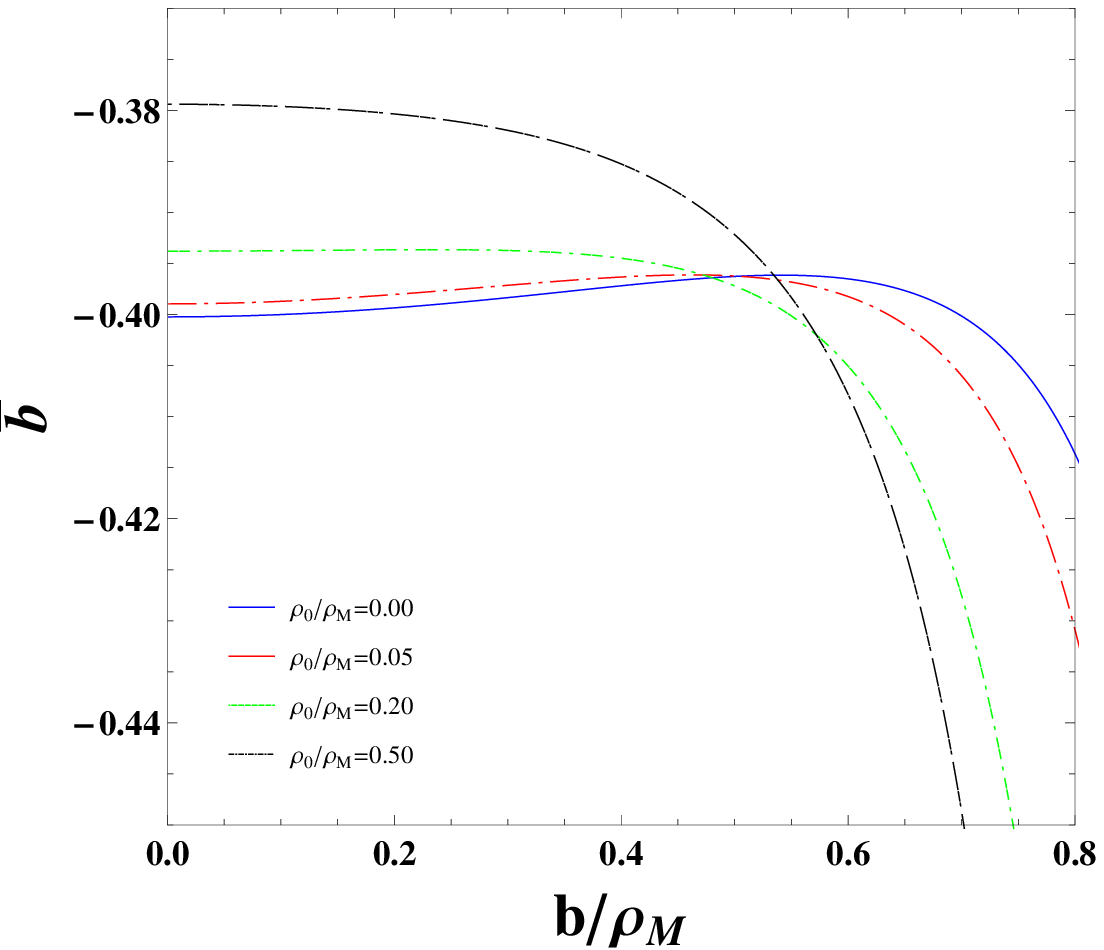}
\caption{Variety of the coefficient $\bar{b}$ with the parameters $\rho_0/\rho_M$ and $b/\rho_M$ in a rotating squashed KK black hole spacetime.}
\end{center}
\end{figure}
with
\begin{eqnarray}
&\bar{a}&=\frac{R(0,\rho_{ps})}{2\sqrt{q(\rho_{ps})}}, \nonumber\\
&\bar{b}&=
-\pi+b_R+\bar{a}\log{\frac{\rho^2_{hs}[C''(\rho_{ps})\mathcal{F}(\rho_{ps})-C(\rho_{ps})\mathcal{F}''(\rho_{ps})]}{u_{ps}
\sqrt{\mathcal{F}^3(\rho_{ps})C(\rho_{ps})}}}, \nonumber\\
&b_R&=I_R(\rho_{ps}), \;\;\;\;\;\;\;\;
u_{ps}=\sqrt{\frac{C(\rho_{ps})}{\mathcal{F}(\rho_{ps})}}.\label{abarbbar}
\end{eqnarray}
Here the quantity $D_{OL}$ is the distance between observer and gravitational lens, $\theta=u/D_{OL}$ is the angular separation between the lens and the image, the subscript ``$ps$" represent the evaluation at $\rho=\rho_{ps}$.  Similarly, one can obtain
the strong gravitational lensing formula for the deflection
angle in the $\psi$ direction ( $\alpha_{\psi}(\theta)$), which has a similar form  with the coefficients differed slightly from $\bar{a}$ and
$\bar{b}$ in Eq.(\ref{abarbbar}). As $\rho_{s}$ tends to $\rho_{ps}$, we
find that the deflection angle $\alpha_{\psi}(\theta)$ also diverges
logarithmically. Since $\alpha_{\psi}(\theta)$ cannot actually be
observed by astronomical experiments, we do not consider it in
the following discussion.
\begin{figure}[ht]
\begin{center}
\includegraphics[width=6cm]{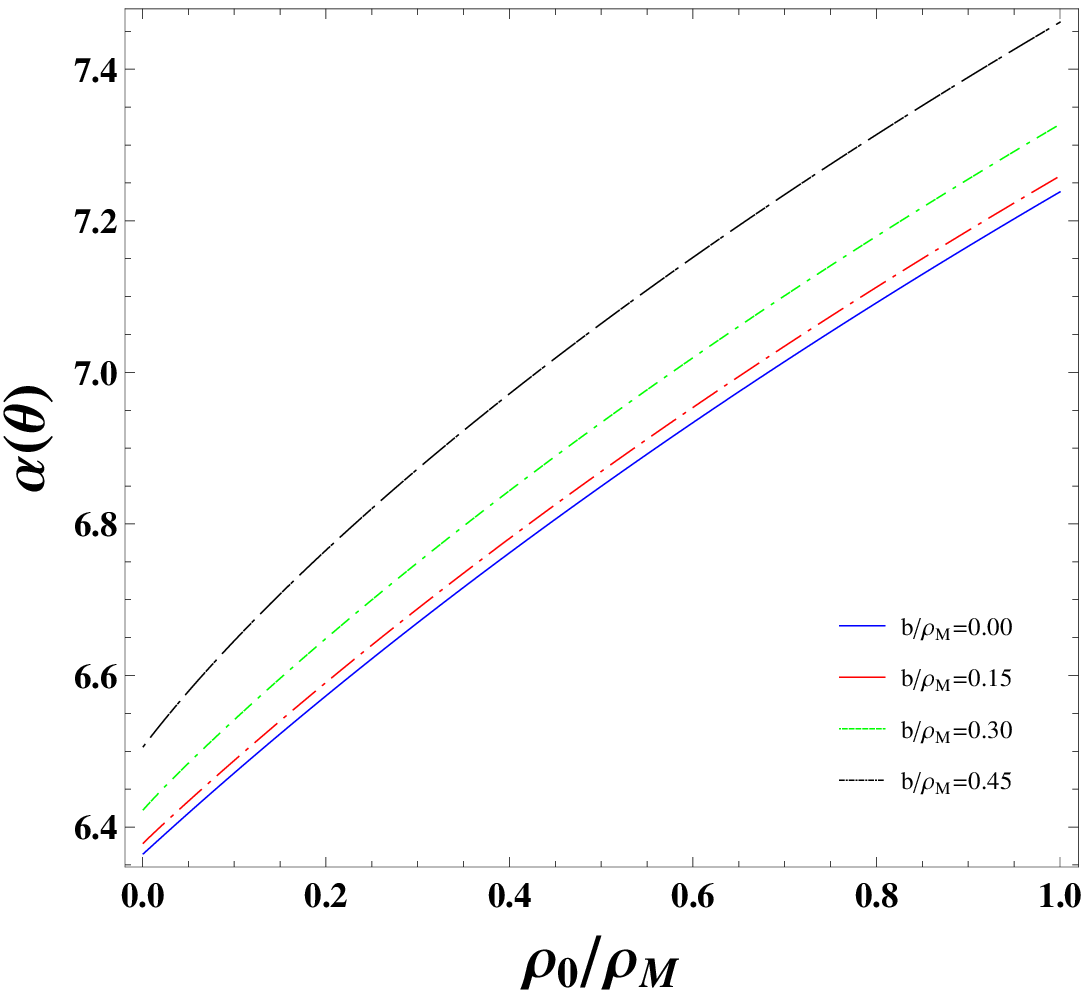}\;\;\;\includegraphics[width=6cm]{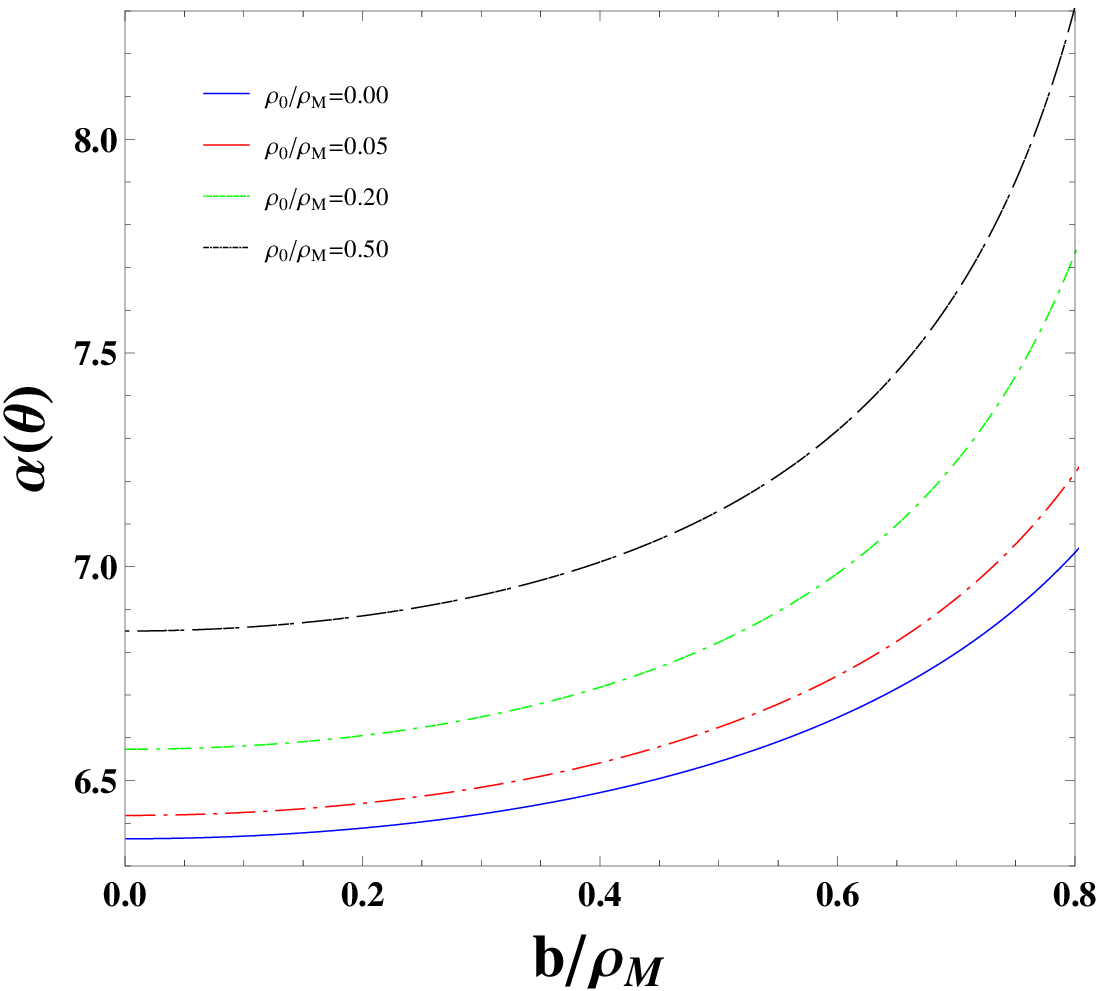}
\caption{Deflection angles in the squashed KK G\"{o}del
black hole spacetime evaluated at $u=u_{ps}+0.003$ as functions of
$\rho_0/\rho_M$ and $b/\rho_M$.}
\end{center}
\end{figure}

We are now in position to probe the properties of strong gravitational lensing in the rotating squashed KK black hole spacetime and explore the effects of the rotation parameter $b$ and the scale of transition $\rho_0$ on the deflection angle in the strong field limit. In the Figs.(2)-(3), we plotted numerically the dependence of the coefficients ( $\bar{a}$ and $\bar{b}$ ) on the parameters $b$ and $\rho_0$. It is shown that the coefficients ( $\bar{a}$ and $\bar{b}$ )  in the strong field limit are functions of the rotation parameter $b$ and the scale of transition $\rho_0$. The coefficient  $\bar{a}$ increases monotonously with $\rho_0$ and $b$. In the extremely squashed case $\rho_0=0$, we find that the change of $\bar{a}$ with the rotation parameter $b$ is different from the change of $\bar{a}$ with the global rotation parameter $j$ in the squashed G\"{o}del KK case in which $\bar{a}$ is independent of $j$ in this extremely squashed case. The variety of $\bar{b}$ with $\rho_0$ and $b$ becomes more complicated. With the increase of $\rho_0$,  $\bar{b}$ increases for the smaller $b$ and decreases for the larger one. With the increase of $b$, $\bar{b}$ first increases and then decreases for the smaller $\rho_0$,  and decreases monotonously for the larger one.
Furthermore, in Fig.(4), we plotted the change of the deflection angle $\alpha(\theta)$ estimated at $u=u_{ps}+0.003$ with $\rho_0$ and $b$, which tells us that in the strong field limit the deflection angles have similar properties of the coefficient $\bar{a}$. This means that the deflection angles of the light rays are dominated by the logarithmic
term in this case.

\section{Observational gravitational lensing parameters}

In this section, we will estimate the numerical values for the observables of gravitational lensing in the strong field limit by assuming that the spacetime of the supermassive black hole at the Galactic center of Milky Way can be described by the rotatiing squashed KK black hole metric (\ref{metricn1}) and
then probe the effects of the rotation parameter $b$ and the scale parameter $\rho_0$ on the observables in the strong gravitational lensing.
As the source and observer are far enough from the lens, the lens equation can be approximated well as \cite{Bozza3}
\begin{eqnarray}
\gamma=\frac{D_{OL}+D_{LS}}{D_{LS}}\theta-\alpha(\theta) \; mod
\;2\pi
\end{eqnarray}
where $\gamma$ is the angle between the direction
of the source and the optical axis. $D_{LS}$ is the lens-source distance and $D_{OL}$ is the observer-lens distance.  $\theta=u/D_{OL}$ is the angular
separation between the lens and the image. Following
ref.\cite{Bozza3},  we here
consider only the case in which the source, lens and observer are
highly aligned. In this simplest case, the angular separation between the lens and the $n-$th relativistic image can be written as
\begin{eqnarray}
\theta_n\simeq\theta^0_n\bigg(1-\frac{u_{ps}e_n(D_{OL}+D_{LS})}{\bar{a}D_{OL}D_{LS}}\bigg),
\end{eqnarray}
with
\begin{eqnarray}
\theta^0_n=\frac{u_{ps}}{D_{OL}}(1+e_n),\;\;\;\;\;\;e_{n}=e^{\frac{\bar{b}+|\gamma|-2\pi
n}{\bar{a}}}.\label{st1}
\end{eqnarray}
Here $\theta^0_n$ is the image position corresponding to
$\alpha=2n\pi$, and $n$ is an integer. As $n\rightarrow \infty$,  one can find from Eqs.(\ref{st1}) that $e_n\rightarrow 0$, which implies that the
minimum impact parameter $u_{ps}$ and the asymptotic position of a
set of images $\theta_{\infty}$ obey a simple form
\begin{eqnarray}
u_{ps}=D_{OL}\theta_{\infty}.\label{ups}
\end{eqnarray}
In order to obtain the coefficients $\bar{a}$ and $\bar{b}$, one needs to separate the outermost image from all the others. Following  refs.\cite{Bozza2,Bozza3},  we consider here the simplest case where only the outermost image $\theta_1$ is resolved as a single image and all the remaining ones are packed together at $\theta_{\infty}$. And then the angular separation between the first image and other ones $s$ and the ratio of the flux from the first image and those from the all other images $\mathcal{R}_0$ can be simplified further as \cite{Bozza2,Bozza3}
\begin{eqnarray}
s&=&\theta_1-\theta_{\infty}=\theta_{\infty}e^{\frac{\bar{b}-2\pi}{\bar{a}}},\nonumber\\
\mathcal{R}_0&=&\frac{\mu_1}{\sum^{\infty}_{n=2}\mu_n}=e^{\frac{2\pi}{\bar{a}}}.\label{sR}
\end{eqnarray}
Therefore, one can estimate the strong deflection limit coefficients $\bar{a}$, $\bar{b}$ and the minimum impact parameter $u_{ps}$ by
measuring these three simple observations $s$, $\mathcal{R}_0$, and $\theta_{\infty}$. Comparing their values with those predicted by the theoretical models, we can extract the characteristics information about the compact object stored in the strong gravitational lensing.
\begin{figure}[ht]
\begin{center}
\includegraphics[width=6cm]{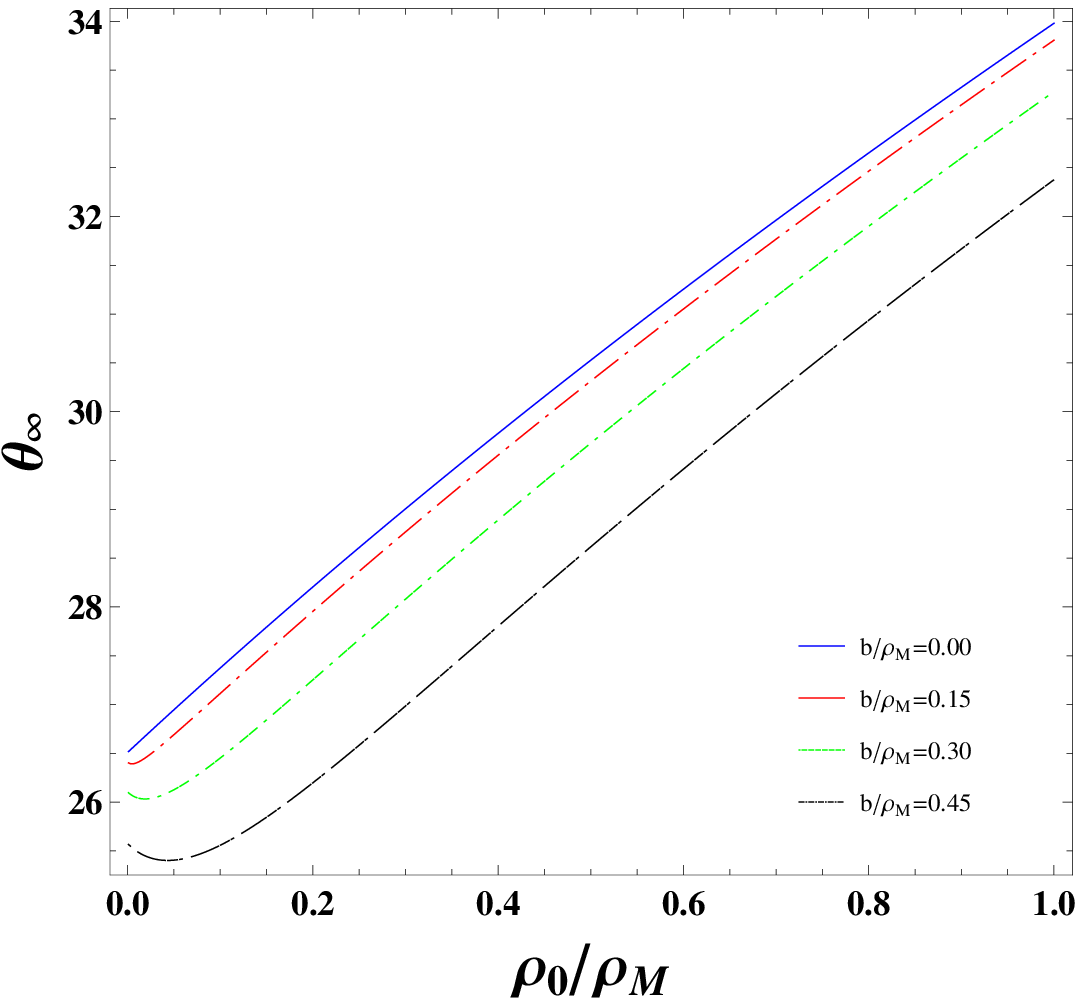}\;\;\;\includegraphics[width=6cm]{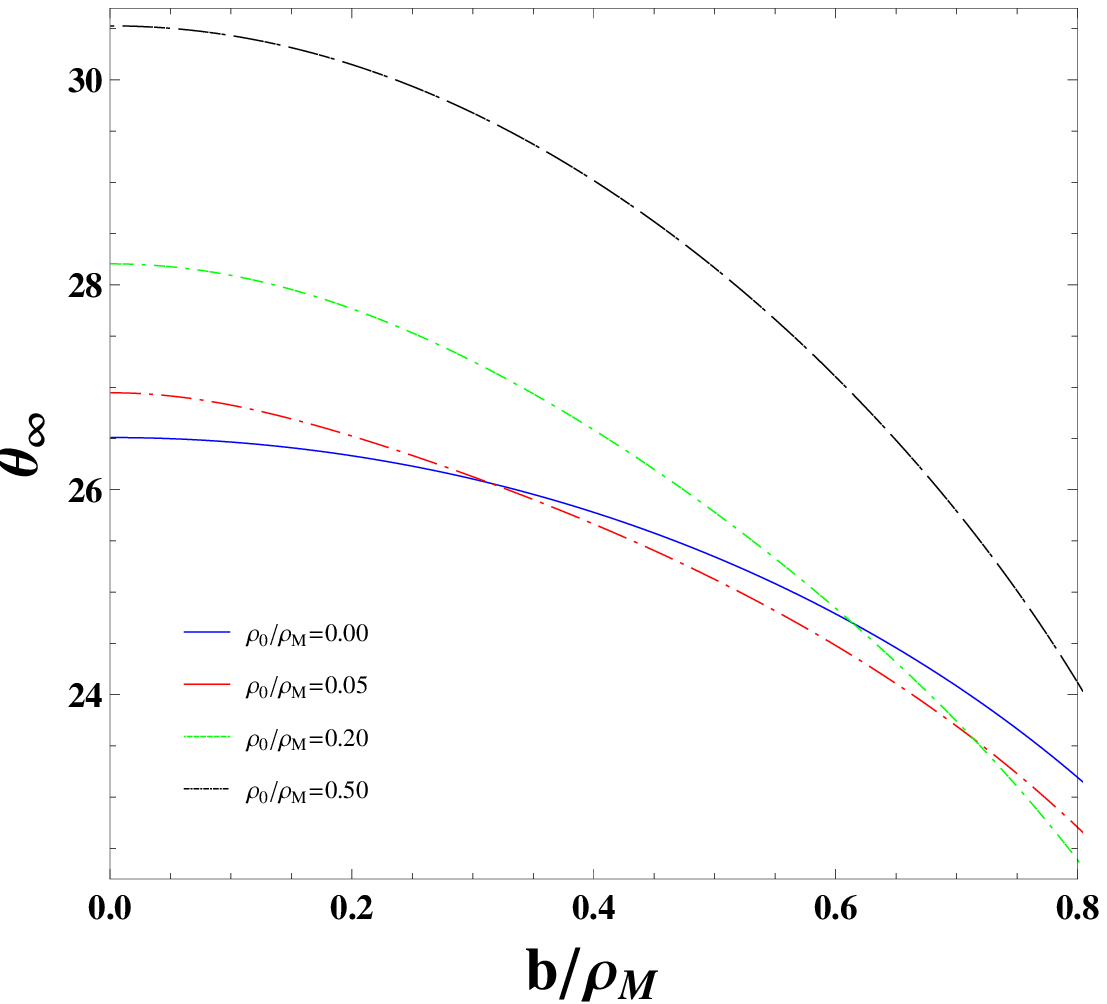}
\caption{Gravitational lensing by the Galactic center black hole.
Variation of the values of the angular position $\theta_{\infty}$
with parameters $\rho_0/\rho_M$ and $b/\rho_M$ in a rotating squashed KK black hole spacetime.}
\end{center}
\end{figure}
\begin{figure}[ht]
\begin{center}
\includegraphics[width=6cm]{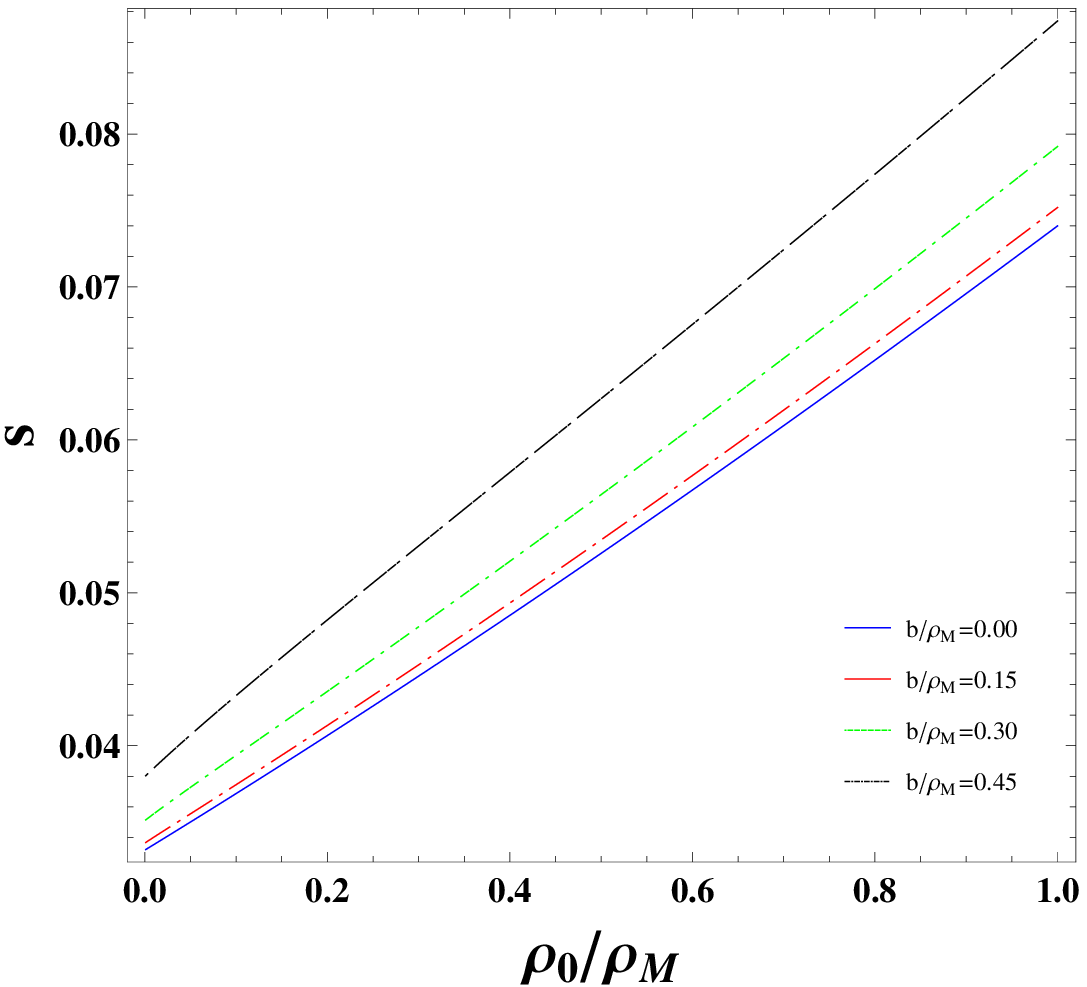}\;\;\;\includegraphics[width=6cm]{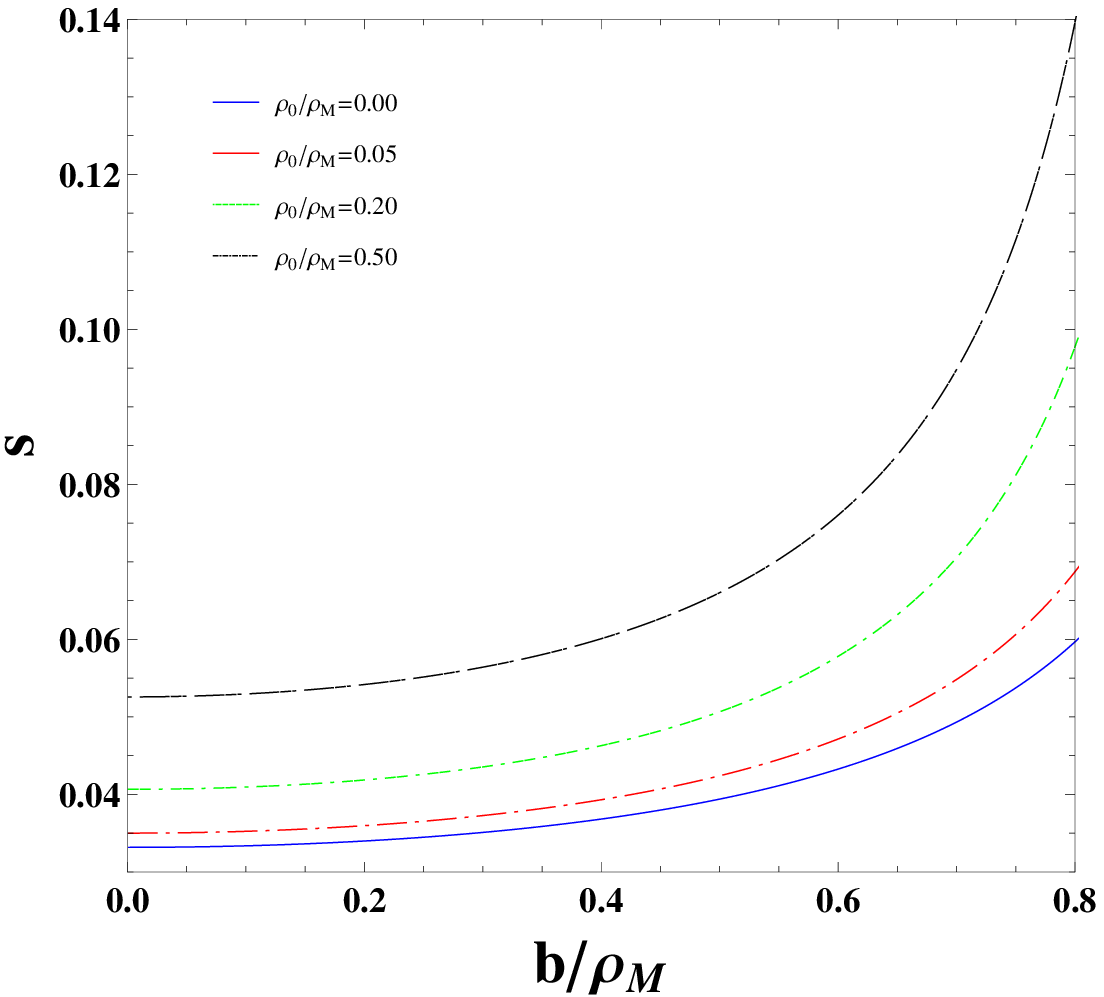}
\caption{Gravitational lensing by the Galactic center black hole.
Variation of the values of the angular separation $s$ with
parameters $\rho_0/\rho_M$ and $b/\rho_M$ in a rotating squashed KK black hole spacetime.}
\end{center}
\end{figure}
\begin{figure}[ht]
\begin{center}
\includegraphics[width=6cm]{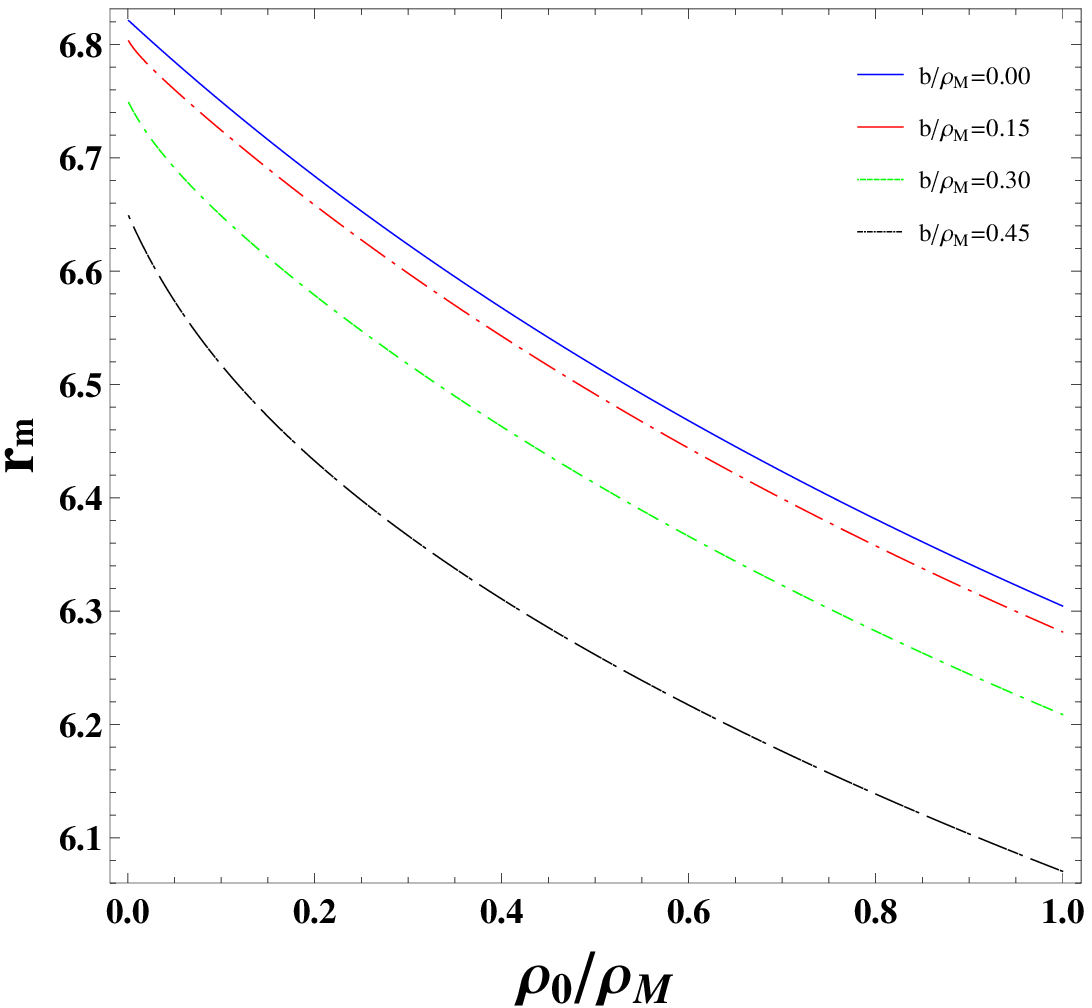}\;\;\;\includegraphics[width=6cm]{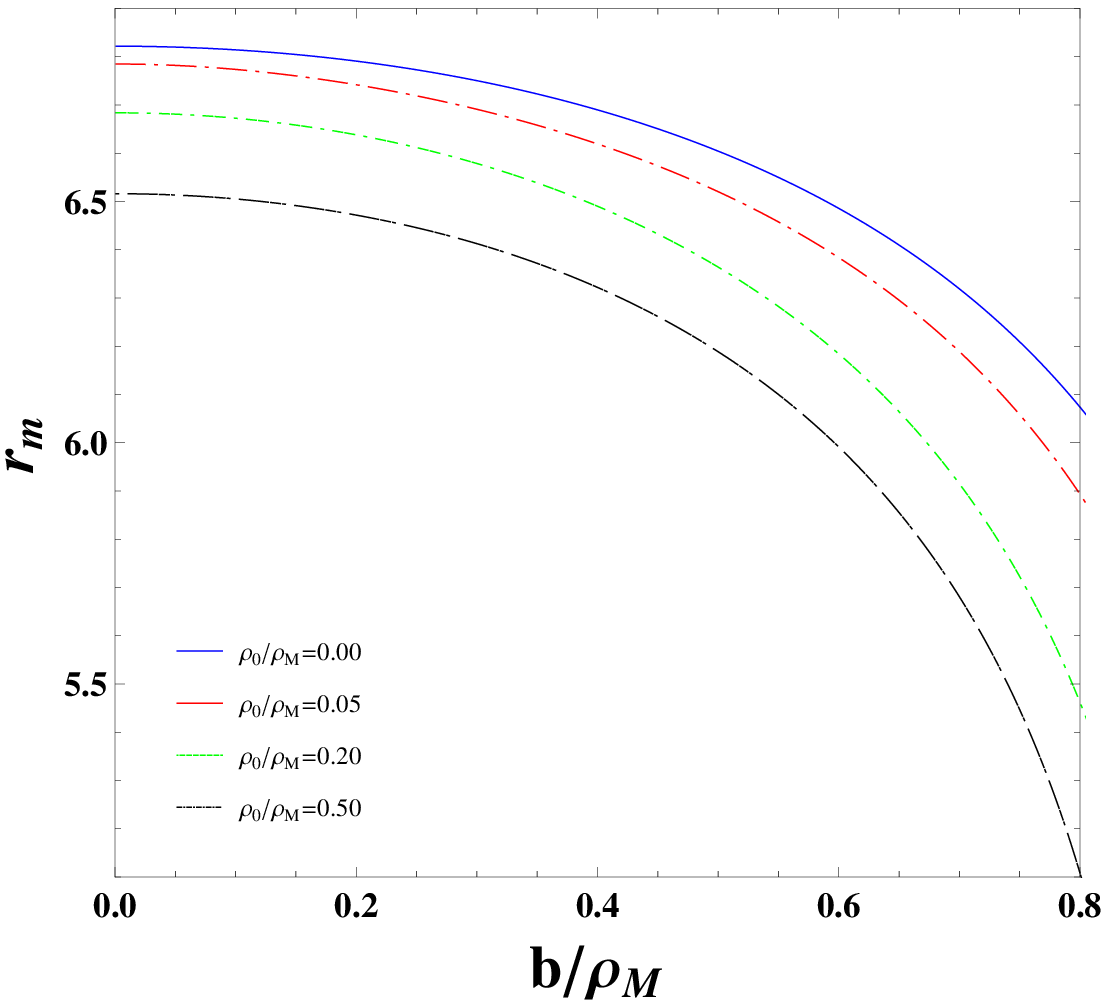}
\caption{Gravitational lensing by the Galactic center black hole.
Variation of the values of the relative magnitudes $r_m$ with
parameters $\rho_0/\rho_M$ and $b/\rho_M$ in a rotating squashed KK black hole spacetime.}
\end{center}
\end{figure}

\begin{table*}[ht]
\begin{center}
\begin{tabular}{c | c | c | c | c}
  \hline
  \hline
    $a/\rho_M$      &    $\rho_0/\rho_M$       &      $\theta_{\infty}$ \,($\mu$ arcsec)   &     $s$\,($\mu$ arcsec)       &   $r_m$ \,(magnitudes)\\

  \hline
 {} & 0.00 & 26.510 & 0.033177 & 6.8219 \\
 {} & 0.05 & 26.947 & 0.035012 & 6.7849 \\
 0.00 & 0.20 & 28.206 & 0.040666 & 6.6838 \\
 {} & 0.50 & 30.528 & 0.052584 & 6.5162 \\
  \hline
 {} & 0.00 & 26.410 & 0.033632 & 6.8047 \\
 {} & 0.05 & 26.691 & 0.035547 & 6.7603 \\
 0.15 & 0.20 & 27.955 & 0.041328 & 6.6583 \\
 {} & 0.50 & 30.315 & 0.053464 & 6.4915 \\
  \hline
 {} & 0.00 & 26.105 & 0.035105 & 6.7505 \\
 {} & 0.05 & 26.127 & 0.037280 & 6.6911 \\
 0.30 & 0.20 & 27.253 & 0.043552 & 6.5788 \\
 {} & 0.50 & 29.678 & 0.056422 & 6.4126 \\
  \hline
 {} & 0.00 & 25.576 & 0.037987 & 6.6507 \\
 {} & 0.05 & 25.406 & 0.040707 & 6.5737 \\
 0.45 & 0.20 & 26.198 & 0.048228 & 6.4326 \\
 {} & 0.50 & 28.616 & 0.062700 & 6.2616 \\
  \hline
\end{tabular}
\caption {Numerical estimation for main observables and the strong field limit coefficients for the black hole at the
center of our Galaxy, which is supposed to be described by a rotating squashed KK black hole spacetime.
$r_m = 2.5 \log \mathcal{R}_0$}
\end{center}
\end{table*}

 Recently, the mass of the central object of our Galaxy is estimated to be $4.4\times 10^6M_{\odot}$  and its distance is around $8.5kpc$ \cite{grf}. This means the ratio of the mass of the central object to the distance $G_4\mathcal{M}/D_{OL} \approx2.4734\times10^{-11}$ . Here $\mathcal{M}$ is the mass of the black hole and
$D_{OL}$ is the distance between the lens and the observer in the $\rho$ coordination rather than that in $r$ coordination because
that in the five-dimensional spacetime the dimension of the black hole mass $\mathcal{M}$ is the square of that in the polar coordination $r$.
Combing with Eqs. (\ref{abarbbar}) and (\ref{sR}), we can estimate the values of the coefficients and observables in strong gravitational lensing in the rotating squashed KK black hole spacetime. In Table I, we list the numerical values of $\theta_{\infty}$, $s$ and $r_{m}$ (which is related to $\mathcal{R}_0$ by $r_m=2.5\log{\mathcal{R}_0}$) for the different values of $b/\rho_M$ and $\rho_0/\rho_M$. Moreover, we also present the dependence of these observables on the parameters $b/\rho_M$ and $\rho_0/\rho_M$  in Figs.(5)-(7).
It is shown that with the increase of the scale $\rho_0$, the angular position of the relativistic images $\theta_\infty$ first decreases and then increases for the non-zero $b$ and increases monotonously in the case with $b=0$. For fixed $\rho_0/\rho_M$, $\theta_\infty$ decreases monotonously with $b$. Moreover, we also find that with the increase of the rotation parameter $b$ and the scale of extra dimension $\rho_0$, the angular separation $s$ increase, while the relative magnitudes $r_m$ decrease monotonously.

Finally, we make a comparison among the strong gravitational lensings in
the rotating squashed KK, the squashed KK G\"{o}del and four-dimensional Kerr black hole spacetimes.
For the photon moving along
the equatorial plane in the four-dimensional Kerr black hole
spacetime \cite{Bozza3,Bozza4}, we know that all of the photon sphere radius, the angular position of the relativistic images
$\theta_{\infty}$ and the relative magnitudes $r_m$ in strong
gravitational lensing decrease with the rotation parameter $a$ of
the black hole, while the coefficient $\bar{a}$, the deflection
angle $\alpha(\theta)$ and the observable $s$ in strong
gravitational lensing increase with the rotation parameter $a$.  These effects of $a$ on the strong gravitational lensing in the Kerr black hole are similar
not only to those of the rotation parameter $b$ in the rotating squashed KK black hole spacetime, but also to those of the rotation parameter
$j$ of cosmological background in
the squashed KK G\"{o}del case, which can be
understandable since all of  $a$, $b$ and $j$ are the rotation
parameters. Therefore, these above features could be looked as the universal effects of the rotation on the strong gravitational lensing.
However, there exist some
distinct properties among the strong gravitational lensings in these three black hole spacetimes. In the rotating squashed KK and the squashed KK G\"{o}del black holes, the photon sphere
radius, the coefficient $\bar{a}$,
$\bar{b}$ and the deflection angle $\alpha(\theta)$ in the $\phi$
direction are independent of whether the photon goes with or against
the rotation of the background.  While in the Kerr
black hole, the values of these quantities for the prograde photons
are different from those for the retrograde photons.
Comparing with the strong gravitational lensings in the rotating squashed KK and the squashed KK G\"{o}del black holes, we find that with increase of $\rho_0$, the coefficient $\bar{a}$, the deflection
angle $\alpha(\theta)$ and the observable $s$ increase and the relative magnitudes $r_m$ decreases in these two kinds of black holes.
Moreover, we find also that there also exist some different effects of $\rho_0$ on the marginally stable orbit radius $\rho_{ps}$ and the angular position of the relativistic images $\theta_{\infty}$. With the increase of $\rho_0$,  the quantities $\rho_{ps}$ and $\theta_{\infty}$ first decreases and then increases in the rotating squashed KK black hole for fixed rotation parameter $b$, but in the squashed
KK G\"{o}del black hole they increase for the smaller global rotation parameter $j$ and decrease for the larger one. In the extremely squashed case $\rho_0=0$, the coefficient $\bar{a}$ in the squashed
KK G\"{o}del black hole is a constant $1$ and is independent
of the global rotation of the G\"{o}del Universe and in the rotating squashed KK black hole it increases monotonously with the rotation parameter $b$.
These information
could help us to understand further the effects of the rotation parameter and the scale of extra dimension on the strong gravitational lensing in the black hole spacetimes.

\section{summary}

We have investigated the strong gravitational lensing in the rotating squashed KK black hole and find that the rotation parameter of black hole and the scale of extra dimension imprint in the radius of the marginally circular photon
orbit, the deflection angle, the coefficients $\bar{a}$, $\bar{b}$ and the observational variables in strong field lensing.
The marginally circular photon radius $\rho_{ps} $, the angular position of the relativistic images
$\theta_{\infty}$ and the relative magnitudes $r_m$ in strong
gravitational lensing decrease with the rotation parameter $b$ of
the black hole, while the coefficient $\bar{a}$, the deflection
angle $\alpha(\theta)$ and the observable $s$ in strong
gravitational lensing increase with the rotation parameter $b$.
With the increase of the extra dimension scale $\rho_0$, the coefficient $\bar{a}$, the deflection angle $\alpha(\theta)$ and the observable $s$ increase, while the relative magnitudes $r_m$ decreases monotonously and
$\rho_{ps} $ first
decreases and then increases in the rotating squashed KK black hole spacetime.
Moreover, we find that in the rotating squashed KK black hole spacetime the marginally circular photon radius $\rho_{ps}$, the coefficient $\bar{a}$,
$\bar{b}$, the deflection angle $\alpha(\theta)$ in the $\phi$
direction and the corresponding observational variables are independent of whether the photon goes with or against
the rotation of the background, which is different with those of in the usual four-dimensional Kerr black hole spacetime. Comparing with the strong gravitational lensings the squashed KK G\"{o}del black holes, we find that there exist some different effects of $\rho_0$ on the marginally circular photon radius $\rho_{ps}$ and the angular position of the relativistic images $\theta_{\infty}$. With the increase of $\rho_0$,  the quantities $\rho_{ps}$ and $\theta_{\infty}$ first decreases and then increases in the rotating squashed KK black hole for fixed rotation parameter $b$, but in the squashed
KK G\"{o}del black hole they increase for the smaller global rotation parameter $j$ and decrease for the larger one. In the extremely squashed case $\rho_0=0$, the coefficient $\bar{a}$ in the squashed
KK G\"{o}del black hole is a constant $1$ and is independent
of the global rotation of the G\"{o}del Universe, but it increases monotonously with the rotation parameter $b$ in the rotating squashed KK black hole.

\begin{acknowledgments}

This work was  partially supported by the National Natural Science Foundation of China under Grant No.11275065, the NCET under Grant
No.10-0165, the PCSIRT under Grant No. IRT0964,  the Hunan Provincial Natural Science Foundation of China (11JJ7001) and the construct
program of key disciplines in Hunan Province. J. Jing's work was
partially supported by the National Natural Science Foundation of
China under Grant Nos. 11175065, 10935013; 973 Program Grant No.
2010CB833004.

\vspace*{0.2cm}
\end{acknowledgments}


\vspace*{0.2cm}
\begin{thebibliography}{99}
\baselineskip=0.6 cm
\bibitem{IM}
H. Ishihara and K. Matsuno, Prog. Theor. Phys. {\bf 116}, 417
(2006).

\bibitem{sq2} S. S. Yazadjiev, Phys. Rev. D {\bf 74},  024022 (2006).

\bibitem{sq3} Y. Brihaye and E. Radu, Phys. Lett. B {\bf 641},
212 (2006).

\bibitem{sq4} H. Ishihara, M. Kimura, K. Matsuno, and S. Tomizawa,
Phys. Rev. D {\bf74}, 047501 (2006); Class. Quantum Grav. {\bf 23},
6919 (2006).

\bibitem{TVN} T. Harmark, V. Niarchos and N. A. Obers,
Class. Quant. Grav. {\bf24}, R1-R90 (2007).

\bibitem{TW} T. Wang, Nucl. Phys. B {\bf 756}, 86 (2006).

\bibitem{sq6} V. Frolov and D. Stojkovic, Phys. Rev. D {\bf 67}, 084004 (2003); H. Nomura, S. Yoshida, M.
Tanabe and K. Maeda, Prog. Theor. Phys. {\bf 114},  707-712 (2005).

\bibitem{SHKT} S. Tomizawa, H. Ishihara, K. Matsuno, and T. Nakagawa, Prog. Theor. Phys. {\bf121}, 823 (2009), arXiv:0803.3873.



\bibitem{sq1hw} H. Ishihara and J. Soda, Phys. Rev. D {\bf76}, 064022 (2007).


\bibitem{sq2hw}  S. Chen, B. Wang and R. Su, Phys. Rev. D {\bf 77}, 024039
(2008).

\bibitem{sqq1}  X. He, B. Wang and S. Chen, Phys. Rev. D {\bf 79}, 084005
(2009);  X. He, S. Chen, B. Wang, R. G. Cai, C. Lin, Phys.Lett.B
{\bf665}, 392 (2008).

\bibitem{sqq2}  H. Ishihara, M. Kimura, R. A. Konoplya, K. Murata, J. Soda and A.
Zhidenko, Phys.Rev.D {\bf77}, 084019 (2008).

\bibitem{Ksq} K. Matsuno and H. Ishihara, Phys. Rev. D {\bf80}, 104037
(2009).


\bibitem{Einstein}A. Einstein, \emph{Science}, \textbf{84},  506 (1936).

\bibitem{schneider}P. Schneider, J. Ehlers, and E. E. Falco "Gravitational
Lenses" \emph{Springer-Verlag, Berlin}, (1992).
\bibitem{Darwin} C. Darwin, Proc. of the Royal Soc. of London {\bf 249}
180 (1959).


\bibitem{Vir} K. S. Virbhadra and G. F. R. Ellis,  Phys. Rev.D {\bf 65}, 103004(2002);  K. S. Virbhadra, D. Narasimha and S. M. Chitre,
Astron. Astrophys. {\bf 337} 1 (1998).  K. S. Virbhadra, G. F. R. Ellis,  Phys. Rev. D {\bf 62} 084003 (2000);  C. M. Claudel, K. S. Virbhadra, G. F. R. Ellis, J. Math. Phys. {\bf 42} 818 (2001).


\bibitem{Fritt} S. Frittelly, T. P. Kling and E. T. Newman, Phys. Rev. D {\bf 61}, 064021 (2000).



\bibitem{Bozza1} V. Bozza, S. Capozziello, G. lovane and G.
Scarpetta, Gen. Rel. and Grav. {\bf 33}, 1535 (2001).

\bibitem{Eirc1}E. F. Eiroa, G. E. Romero and D. F. Torres, Phys. Rev. D {\bf 66},
024010 (2002); E. F. Eiroa, Phys. Rev. D {\bf 71}, 083010 (2005); E.
F. Eiroa, Phys. Rev. D {\bf 73}, 043002 (2006).

\bibitem{whisk} R. Whisker, Phys. Rev. D {\bf71}, 064004 (2005).

\bibitem{Bozza2} V. Bozza, Phys. Rev. D {\bf 66}, 103001 (2002).

\bibitem{Bozza3} V. Bozza, Phys. Rev. D {\bf 67}, 103006 (2003)

\bibitem{Bozza4} V. Bozza, F. De Luca, G. Scarpetta, and M. Sereno, Phys. Rev. D {\bf 72}, 08300 (2005); V. Bozza, F. De Luca, and G. Scarpetta, Phys.
Rev. D {\bf 74}, 063001 (2006).

\bibitem{Gyulchev} G. N. Gyulchev and S. S. Yazadjiev, Phys. Rev. D {\bf75}
023006 (2007);  G. N. Gyulchev and S. S. Yazadjiev, Phys. Rev. D {\bf78}
083004 (2008).


\bibitem{Bhad1} A. Bhadra, Phys. Rev. D {\bf 67}, 103009 (2003).

\bibitem{TSa1} T. Ghosh and S. Sengupta, Phys. Rev. D {\bf 81}, 044013 (2010), arXiv: 1001.5129.

\bibitem{AnAv}A. N. Aliev and P. Talazan, Phys. Rev. D {\bf80}, 044023
(2009), arXiv:0906.1465.


\bibitem{gr1} C. Ding, C. Liu, Y. Xiao, L. Jiang and R. Cai,
 Phys. Rev. D {\bf88} 104007 (2013); E. F. Eiroa and C. M. Sendra, Phys. Rev. D {\bf88}, 103007 (2013).

\bibitem{gr2} S. Chen and J. Jing, Phys. Rev. D {\bf85}, 124029 (2012).

\bibitem{gr3} S. Wei, Y. Liu, C. Fu and K. Yang, JCAP {\bf1210},  053 (2012); S. Wei, Y. Liu, Phys. Rev. D {\bf85}, 064044 (2012).


\bibitem{Kraniotis} G. V. Kraniotis, Class. Quant. Grav. {\bf28}, 085021
(2011).


\bibitem{schen} Y. Liu, S. Chen and J. Jing, Phys. Rev. D {\bf81},124017
(2010).

\bibitem{schen2} S. Chen, Y. Liu and J. Jing, Phys. Rev. D {\bf83}, 124019
(2011)

\bibitem{JH} J. Sadeghi, H. Vaez, Phys. Lett. B {\bf728}, 170-182 (2014), arXiv:1310.4486.

\bibitem{JAH} J. Sadeghi, A. Banijamali and H. Vaez, arXiv:1205.0805.

\bibitem{Komar} A. Komar, Phys. Rev. {\bf113}, 934 (1959);  Phys. Rev. {\bf 129}, 1873 (1963).
\bibitem{Komar1} R. M. Wald, \textit{General Relativity}, ( University of Chicago Press, Chicago, 1984).

\bibitem{grf} R. Genzel, F. Eisenhauer and S. Gillessen, Rev. Mod. Phys. {\bf82}, 3121 (2010); arXiv:1006.0064.

\end{thebibliography}
\end{document}